\documentclass[%
 reprint,
 amsmath,amssymb,
 aps,prb,
]{revtex4-2}

\usepackage{amsmath}
\usepackage{xcolor}
\usepackage{makecell}
\usepackage[mathscr]{euscript}
\usepackage{graphicx}
\usepackage{dcolumn}
\usepackage{bm}
\usepackage{hyperref}

\begin{document}



\title{{\em Ab initio} Investigation of Thermal Transport in Insulators: Unveiling the Roles of Phonon Renormalization and Higher-Order Anharmonicity}

\author{Soham Mandal}
 
\author{Manish Jain}%
\affiliation{Centre for Condensed Matter Theory, Department of Physics, Indian Institute of Science, Bangalore-560012, India}

\author{Prabal K. Maiti}
\email{maiti@iisc.ac.in}
\affiliation{Centre for Condensed Matter Theory, Department of Physics, Indian Institute of Science, Bangalore-560012, India}

\date{\today}

\begin{abstract}
The occurrence of thermal transport phenomena is widespread, exerting a pivotal influence on the functionality of diverse electronic and thermo-electric energy-conversion devices.
The traditional first-principles theory governing the thermal and thermodynamic characteristics of insulators relies on the perturbative treatment of interatomic potential and ad-hoc displacement of atoms within supercells. 
However, the limitations of these approaches for highly anharmonic and weakly bonded materials, along with discrepancies arising from not considering explicit finite temperature effects, highlight the necessity for a well-defined quasiparticle approach to the lattice vibrations.
To address these limitations, we present a comprehensive numerical framework in this study, designed to compute the thermal and thermodynamic characteristics of crystalline semiconductors and insulators.
The self-consistent phonon renormalization method we have devised reveals phonons as quasiparticles, diverging from their conventional characterization as bare normal modes of lattice vibration.
The extension of the renormalization impact to interatomic force constants (IFCs) of third and fourth orders is also integrated and demonstrated.
For the comprehensive physical insights, we employed an iterative solution of the Peierls-Boltzmann transport equation (PBTE) to determine thermal conductivity and carry out Helmholtz free energy calculations, encompassing anharmonicity effects up to the fourth order.
In this study, we utilize our numerical framework to showcase its applicability through an examination of phonon dispersion, phonon linewidth, anharmonic phonon scattering, and temperature-dependent lattice thermal conductivity in both highly anharmonic materials (NaCl and AgI) and weakly anharmonic materials (cBN and 3C-SiC).
Our study finds that ignoring higher-order (4-phonon) phonon scattering processes is not viable for materials with strong anharmonicity in their interatomic potential. Meanwhile, renormalization demonstrates negligible impact on materials characterized by weak anharmonicity.
The investigation also explores the effect of fourth-order Helmholtz free energy and pressure-dependent thermal conductivity of NaCl crystal.
The theoretical framework developed in this work will help us to understand the physical insight of phonon and phonon-driven thermal and thermodynamic properties of materials and offer valuable guidance for the strategic development of efficient thermal management techniques.
\end{abstract}

\maketitle


\section{\label{sec:Introduction} INTRODUCTION}

Lattice vibrations are the primary heat carriers in crystalline insulators and semiconductors. 
The quasi-particles of quantized lattice vibration is known as a phonon. 
Understanding phonon and phonon-driven properties in materials are of immense importance in the domain of material science, solid-state physics, and chemistry. 
This is due to the evergrowing technological application of such materials. 
Phonons play a role in several physical phenomena, including thermal transport, thermal expansion of solids, the stability of crystalline solids, and structural phase transitions, either directly or indirectly.
Phonons and several phonon-assisted physical phenomena have been theoretically analyzed and well-established for more than a century with their experimental validations. 
Rapid increase of high performance computational power has enhanced our ability to use first-principle computation to predict phonon-driven physical phenomena through numerical techniques.
These traditional {\em ab initio} schemes range from simple models taking into account only the harmonic approximation \cite{phon_FD, phonopyTOGO2015} to predict phonon dispersions, the density of states, and heat capacity to more sophisticated approaches to include anharmonicity in the interatomic potential to predict thermal expansion and thermal conductivity of materials\cite{phono3py_2015, ShengBTE_2014}. 
These harmonic and quasi-harmonic approximations are excellent in capturing the aforementioned phonon-driven properties of weakly anharmonic strong covalently bonded materials, like silicon and diamond\cite{QHA}. 

In the case of materials with a significant degree of anharmonicity in their interatomic potentials, such as alkali halides and heavy-metal chalcogenides, simple models of calculating the normal modes of lattice vibration are inadequate.
Materials with extreme anharmonicities, such as solid helium, typically have bare phonon frequencies having imaginary values\cite{Wallace}.
Renormalization makes these squared frequencies positive as anharmonicity is sufficiently large.
Furthermore, at a higher temperature (depending on the Debye temperature of a specific material), conventional ab initio approaches fail to comply with experimental measurements.
Therefore, a well-defined and robust quasi-particle methodology is crucial to cover weakly bonded, highly anharmonic materials and to fit measured data over the entire experimental temperature ranges.

The interatomic force constants (IFCs) calculation in traditional {\em ab initio} approaches involves the density-functional perturbation (DFPT)\cite{DFPT, abinit_DFPT} method or the finite-difference supercell\cite{IFC_symm, phonopy-phono3py-JPCM, phonopy-phono3py-JPSJ, phono3py_2015} method.
The determination of IFCs in DFPT relies on analyzing the analytical derivatives of the potential energy.
It can efficiently determine harmonic IFCs and related properties of materials.
While the utilization of DFPT has been reported for calculating third-order anharmonic IFCs\cite{D3Q}, the computational cost of determining higher-order anharmonic IFCs is generally prohibitive. 
Furthermore, DFPT does not explicitly incorporate finite temperature effects.
The finite-difference supercell method approximates the derivative of the potential energy by employing finite differences, and by calculating the forces on atoms within the perturbed supercell at 0K, the IFCs can be determined.
The perturbation is accomplished by ad hoc displacement of atoms within the supercell.
To compute higher-order anharmonic IFCs (beyond 3rd order), force calculation for a considerable number of perturbed snapshots is necessary for the finite-difference supercell method\cite{FourPhonon}.
Recent studies\cite{Ruan4ph_2016, Ruan4ph_2017, Ruan_4ph, 4-ph_KPal} have explored the incorporation of fourth-order IFCs and four-phonon scattering processes across various materials to reduce discrepancies between theoretical predictions and experimental observations of the thermal transport properties.
Classical force field (FF) based molecular dynamics-based approaches for thermal conductivity calculation, using Green-Kubo response function-based formalism\cite{GK_prl, GK_prb} or the non-equilibrium heating-cooling method\cite{PCCP} can take into account finite temperature effects and the inherent anharmonicity.
In our current work, we will employ the temperature-dependent effective potential (TDEP)\cite{TDEP_2011, TDEP_3rd} method, which is a computationally efficient framework capable of addressing the aforementioned challenges. 
TDEP has been used in recent studies to determine the phonons and associated physical properties of materials\cite{a-TDEP, AnkitJain_2020, alamode_2014}.

In this work, we have developed a unified numerical scheme to calculate IFCs up to the fourth order utilizing the temperature-dependent force-displacement data from {\em ab initio} molecular dynamics (AIMD). 
We have also developed a Thermal Stochastic Snapshot (TSS) technique to avoid the computational overhead of AIMD.
Our current study also involves a numerical approach for the self-consistent phonon renormalization and calculation of thermal transport properties of the insulating crystalline solids. 
In this renormalization method, bare second-order IFCs are renormalized only up to first-order using the statistical operator renormalization technique. 
This renormalization process can also be realized through the one-phonon propagator approach in the many-body theory\cite{GDMahan}.
In the results section, we will demonstrate the phonon-driven physical properties of materials with pronounced anharmonicity (NaCl and AgI) and those with weak anharmonicity (cBN and 3C-SiC) in their interatomic potentials. 
This work thoroughly investigates the impact of self-consistent renormalization of harmonic and anharmonic IFCs and higher-order phonon scattering processes on thermal transport properties.
Finally, our investigation will examine the influence of fourth-order Helmholtz free energy on the lattice constant and pressure-dependent thermal conductivity of crystalline NaCl.
\section{\label{sec:Methods} METHODS}
The potential energy of a crystal due to the interaction among the constituent atoms, $U$, can be expanded in the displacements from the initial configuration, assuming $U$ is an analytic function of the positions of the atoms, as:
\begin{widetext}
\begin{equation}\label{eq:PE}
\begin{aligned}
	    & U = U_0 + \sum\limits_{{N_1 \mu}} 
	        \sum\limits_{\alpha}  
	        \Phi_1^{\alpha }({N_1 \mu}) u_{\alpha}(N_1 \mu) + 
            \sum\limits_{{N_1 \mu},{N_2 \nu}} 
	        \sum\limits_{\alpha,\beta} \frac{1}{2!} 
	        \Phi_2^{\alpha \beta}({N_1 \mu}, {N_2 \nu}) 
	        u_{\alpha}(N_1 \mu) u_{\beta}(N_2 \nu) + \\
	    &    \sum\limits_{{N_1 \mu},{N_2 \nu},{N_3 \pi}}
	        \sum\limits_{\alpha,\beta,\gamma} {\frac{1}{3!}} 
	        \Phi_3^{\alpha \beta \gamma}({N_1 \mu}, {N_2 \nu}, {N_3 \pi})
	        u_{\alpha}(N_1 \mu) u_{\beta}(N_2 \nu) u_{\gamma}(N_3 \pi) + \\
	    &    \sum\limits_{{N_1 \mu},{N_2 \nu},{N_3 \pi},{N_4 \rho}}
	        \sum\limits_{\alpha,\beta,\gamma,\delta} \frac{1}{4!} 
	        \Phi_4^{\alpha \beta \gamma \delta}
	        ({N_1 \mu}, {N_2 \nu}, {N_3 \pi}, {N_4 \rho}) 
	        u_{\alpha}(N_1 \mu) u_{\beta}(N_2 \nu) 
	        u_{\gamma}(N_3 \pi) u_{\delta}(N_4 \rho) + \ldots
\end{aligned}
\end{equation}
\end{widetext}

Here $U_0$ is the potential energy at the equilibrium. $\Phi_1$,  $\Phi_2$,  $\Phi_3$, and  $\Phi_4$ are first-, second-, third-, and fourth-order interatomic force constants (IFCs), respectively. $N_1$, $N_2$, $N_3$, and $N_4$ denote the lattice points of the crystal. $\mu$, $\nu$, $\pi$, and $\rho$ run over the basis atoms; $\alpha$, $\beta$, $\gamma$, and $\delta$ are cartesian indices. The first-order derivative of crystal potential $U$ gives the force on atoms: 
\begin{equation}\label{eq:Force}
\begin{aligned}
    & F^{\alpha}({ N_1 \mu}) = - \frac{\partial U}{\partial u_{\alpha}(N_1 \mu)}
\end{aligned}
\end{equation}

\begin{table*} [!hbt]
\begin{ruledtabular}
{\renewcommand{\arraystretch}{1.2}
    \begin{tabular}{|c|l|l|l|l|l|}
        Material & Order of IFC & Nearest neighbour & Cutoff distance($\AA$) & Total IFCs & Independent IFCs\\ \hline
             & 2nd & 11 & 9.88 & 2790 & 55 \\ \cline{2-6}
        NaCl & 3rd & 5 & 6.55 & 175446 & 1240 \\ \cline{2-6}
             & 4th & 2 & 4.40 & 1111158 & 2627 \\ \hline
             & 2nd & 11 & 10.75/6.10/7.43 & 2700 & 74 \\ \cline{2-6}
        AgI/cBN/3C-SiC & 3rd & 6 & 8.05/4.56/5.56 & 272214 & 3860 \\ \cline{2-6}
             & 4th & 2 & 4.90/2.79/3.39 & 795906 & 3758 \\ 
    \end{tabular}
}
    \caption{\label{Tab:IFC_CutOff}Number of independent IFCs up to 4th order after applying space group symmetry, permutation symmetry, and acoustic sum rule of the IFCs.}
\end{ruledtabular}
\end{table*}
Our calculation of the interatomic force constants (up to the fourth order) involved conducting {\em ab initio} molecular dynamics (AIMD) simulations at a designated temperature on a supercell of the crystal under consideration. 
Unless explicitly stated otherwise, all MD simulations conducted in this study utilize a $5\times5\times5$ supercell configuration.
MD trajectories provide us with the displacements ($u_{\alpha}(N_1 \mu)$) and forces ($F^{\alpha}({ N_1 \mu})$) on each atom in Eq. \ref{eq:Force} at every time step. 
Eq. \ref{eq:Force} can be rewritten as a matrix equation, $\mathcal{D}x = f$, with these displacement and force datasets. Here, the matrix elements of $\mathcal{D}$ are constructed from the atomic displacements; $f$ is a row of atomic forces; $x$ is the row of IFCs to calculate. The matrix equation is usually an overdetermined system. 
A least-squares technique is used to calculate the IFCs by inverting the displacement matrix $\mathcal{D}$ through the Moore-Penrose pseudoinverse\cite{laug}.

\subsection{\label{sec:Symmetry}Symmetry}
In general, the number of IFCs for the $d$-th order is $(3N)^d$, where $N$ is the total number of atoms in the crystal. As such, solving the matrix equations for determining higher order IFCs (beyond fourth order) is computationally prohibitive. To date even calculations including fourth-order IFCs have only been attempted for simplest materials like Silicon and NaCl. We make several approximations to make these calculations tractable. While in principle, one should include all higher-order terms in Eq. \ref{eq:PE}, we include the terms till fourth order. To further reduce the number of IFCs, we use cutoff distances, as the third-, fourth- and non-polar contribution of the second-order IFCs fall rapidly with distance. 
Ewald-Summation technique is used to separate the long-range part of second-order IFCs caused by dipole-dipole interactions in polar materials, as explained in a later section. 
This allows us to identify a suitable set of IFCs for a specific temperature that can accurately match the forces obtained from molecular dynamics simulations.
We reduce the computational cost further by exploiting permutation symmetry, point-group symmetry, and acoustic sum rule (ASR) of the IFCs. The specifics of these symmetries are discussed in Ref\cite{IFC_symm}. 
These symmetries provide constrained equations among the IFCs, from which we extract a set of independent IFCs. 
After determining the independent IFCs, the rest of the IFCs can be reconstructed as their linear combinations. We reconstruct the whole set of IFCs for the calculation of material properties. For determining point-group symmetries, 
we employ the SPGLIB\cite{Spglib} crystal-symmetry library to find the transformation matrices of the point group operations. 
The number of independent IFCs drastically reduces when we combine these symmetries, as illustrated in Table \ref{Tab:IFC_CutOff}, making it more computationally feasible to calculate up to the fourth order. 
\subsection{\label{sec:Long-range}Long-range correction}
The vibration-induced dipole-dipole interaction contribution to the second-order IFCs diminishes at a rate of $\sim 1/r^{3}$ with distance. 
It poses a problem in the case of polar materials, as we cannot set a cutoff for second-order IFCs and, in principle, must extend them to infinity. 
Most of the available packages address this issue by incorporating a non-analytic component into the Dynamical matrix\cite{phonopyTOGO2015, ShengBTE_2014}.
In this mixed-space approach\cite{mixed_space_2010}, the short-range second-order IFCs are calculated in real space separately.
To incorporate long-range correction, a direction-dependent non-analytical term (in reciprocal space) is introduced to the Dynamical matrix.
This correction is applicable only for the long wavelength $(q \to 0)$ limit. 
With a reasonable level of accuracy, this method illustrates the LO-TO splitting of polar materials along the high symmetry path. 
In our case, we used the Ewald Summation technique, as outlined in Ref.\cite{EwaldNonAnalytic, EwaldNonAnalytic2}, to compute the harmonic IFCs in reciprocal space from the Born effective charge and dielectric tensor of the material.
We then use an inverse Fourier transform to convert the reciprocal space IFCs resulting from the dipole-dipole interaction into real-space IFCs. 
The long-range contribution to the force is then calculated by multiplying these newly created real-space IFCs with the displacement dataset already obtained from MD.
We eliminated this long-range force component from the MD force to fit only the short-range component of the IFCs. 
Subsequently, the harmonic IFCs (in reciprocal space) obtained through the Ewald Summation technique, which includes the ionic contribution, are incorporated into the Dynamical matrix to accurately determine the phonon dispersion, group velocity, and density of states. 
Therefore, the complete dynamical matrix is defined as:
\begin{eqnarray}\label{eq:DynMat}
    D^{\alpha \beta}(\mu \nu, q) = \frac{1}{\sqrt{M_\mu M_\nu}}\sum_{N} \Phi_{2(short-range)}^{\alpha \beta}(0\mu, N\nu) e^{i \mathbf{q} \cdot R(N)} \nonumber\\ +\: Ewald\: term.\nonumber\\
\end{eqnarray}
where $M_\mu$ is the mass of the basis atom type $\mu$. The phonon frequency ($\omega_{qs}$) and eigenvector ($\epsilon (qs)$) are obtained by performing diagonalization of the dynamical matrix.

\subsection{\label{sec:TSS}Thermal Stochastic Snapshots}
The method we discussed is based on the Temperature-dependent Effective Potential (TDEP) method described by Hellman et al. \cite{TDEP_2011}
TDEP requires expensive {\em ab initio} molecular dynamics (AIMD) on a supercell at a finite temperature to map it into the model potential energy given by Eq. \ref{eq:PE}. 
AIMD is cumbersome computationally as it requires a sufficiently long run to achieve equilibrium at a specific temperature, followed by the selection of well-spaced snapshots for proper sampling of the Born-Oppenheimer (BO) surface of the supercell. 
To circumvent the computation-intensive nature of this method while still retaining all finite temperature effects, we opt for the thermal stochastic snapshot (TSS) technique, as described in Ref\cite{TSS_2006, TDEP_TSS, Ravichandran_2018}. 
In the TSS scheme, the atoms $(Nv)$ in the supercell are displaced following the equation:
\begin{eqnarray}\label{eq:TSS}
    u_\alpha (N \nu) = \sqrt{\frac{\hbar}{M_\nu N_0}} \sum_{qs} \sqrt{\frac{2n_{qs}+1}{\omega_{qs}}} \cos(2\pi \xi_{(1,qs)}) \nonumber\\
    \times \sqrt{-\ln(1-\xi_{(2,qs)})} \epsilon_{\alpha} (qs, \nu) e^{i q \cdot R(N)}
\end{eqnarray}
Where $\xi_{(1, qs)}$ and $\xi_{(2, qs)}$ are two mode-dependent random numbers generated from a uniform distribution between 0 and 1, with $(qs)$ referring to phonon modes. 
To ensure that the displacement $u_\alpha (N \nu)$ is a real value, the random numbers are restricted such that $\xi_{(1, qs)} = \xi_{(1, -qs)}$ and $\xi_{(2, qs)} = \xi_{(2, -qs)}$. 
$\omega_{qs}$ and $\epsilon_\alpha (qs, \nu)$ are the frequency and eigenvector of the phonon mode $(qs)$, respectively. 
$N_0$ is the number of q-points in the Brillouin zone (BZ) sampled. 
$n_{qs}$ is the Bose-Einstein distribution corresponding to the temperature $T$. 
The displacement is consistent with the canonical ensemble, which takes into account complete quantum statistics, thereby including all finite temperature effects. 
Additionally, it also accounts for the zero-point motions of atoms, which is not achievable in AIMD. 
These supercells of displaced atoms are input into an {\em ab initio} force calculator to calculate the force on each atom. We employed the SIESTA\cite{SIESTA_2002} package to carry out the force computation at the Gamma point for each stochastically perturbed supercell.
A limitation of utilizing Eq. \ref{eq:TSS} is that it necessitates prior knowledge of phonon frequencies ($\omega_{qs}$)and eigenvectors ($\epsilon_\alpha (qs, \nu)$), which are the calculations we aim to perform. 
As such, a self-consistent approach is utilized.
The procedure is initiated using the IFCs obtained through TDEP by conducting AIMD at a very low temperature (80K for NaCl, AgI, cBN and 3C-SiC studied here). We initiate TSS calculations from the TDEP-generated IFCs to obtain an initial set of IFCs.
By using these IFCs as feedback, we generate a new set of a force-displacement dataset through TSS and obtain a new set of IFCs. 
This procedure is repeated until the desired accuracy is achieved self consistently. 
To determine the IFCs of the next higher temperature, we utilized the IFCs from the previous temperature and subsequently executed the self-consistent cycle. 
By doing so, we only had to run AIMD once to initiate the process.
\subsection{\label{sec:Ren}Renormalization}
In the self-consistent phonon renormalization approach, an effective harmonic IFC is obtained by combining unrenormalized (bare) second-order IFCs and fourth-order anharmonic IFCs. 
The statistical operator renormalization technique produces the equation for the renormalized harmonic IFC, which is stated as follows\cite{Wallace, Ravichandran_2018, Ren_Kpal}:
\begin{eqnarray}\label{eq:Renorm}
     \Psi_2^{\alpha \beta}(0\mu, N\nu) = \Phi_2^{\alpha \beta}(0\mu, N\nu) + \nonumber\\
     \frac{\hbar}{4 N_0} \sum_{N_3 \pi} \sum_{N_4 \rho} \sum_{\gamma \delta} \sum_{q s} \Phi_4^{\alpha \beta \gamma \delta}(0\mu, N\nu, N_3 \pi, N_4 \rho)  \nonumber\\ 
    \times e^{i \mathbf{q} \cdot (R(N_3) - R(N_4))} E_{\gamma}(qs, \rho) E_{\delta}(qs, \pi) \frac{2 n_{qs} + 1}{\Omega_{qs} \sqrt{M_\rho M_\pi}}
\end{eqnarray}

Where $\Phi_2^{\alpha \beta}(0\mu, N\nu)$  and $\Phi_4^{\alpha \beta \gamma \delta}(0\mu, N\nu, N_3 \rho, N_4 \pi)$ stands for bare second- and fourth-order IFCs, and $\Psi_2^{\alpha \beta}(0\mu, N\nu)$ is the renormalized second-order IFCs. 
The renormalization process can also be realized by utilizing the one-phonon propagator approach\cite{GDMahan}.
The expression mentioned above is obtained by truncating an infinite series that includes contributions from all even-order IFCs.
As the statistical average is calculated over the effective harmonic phonon states, $\Omega_{qs}$ and $E_{\alpha}(qs, \nu)$ in Eq. \ref{eq:Renorm}, represent the eigenvalue and eigenvector of renormalized phonon, which are unknown initially. 
As such, the problem is solved self-consistently. 
We start the self-consistent loop from the eigenvalue and eigenvectors of bare second-order IFCs ($\Phi_2^{\alpha \beta}(0\mu, N\nu)$). 
During the following iterations, the second term in Eq. \ref{eq:Renorm} is revised using the eigenvalues and eigenvectors of the renormalized harmonic IFCs ($\Psi_2^{\alpha \beta}(0\mu, N\nu)$) from the previous iteration.
The loop is repeated until the desired level of precision is achieved.
It is important to note that, thus far, no interaction among the phonon states has been considered.
The present study does not delve into the impact of the interaction among self-consistent phonon states in the energy levels, mediated by the odd terms in the crystal potential, as it is outside the scope of this work. 
Following the procedure described in Ref.\cite{Ravichandran_2018}, we extend the renormalization effect to anharmonic IFCs of third and fourth order. 
The initial step to achieve this involves the multiplication of recently constructed renormalized second-order IFCs ($\Psi_2^{\alpha \beta}(0\mu, N\nu)$) with the available displacement dataset ($u_\alpha (N \nu)$) from TSS (or AIMD), which enables the calculation of the force ($f_{ren.}^{2nd}$) attributed to the renormalized harmonic IFCs. 
Afterward, we subtract the calculated force from the actual {\em ab initio} force ($f_{\textit{abinitio}}$), which has already been obtained from TSS (or AIMD).
The rest, which remains in Eq. \ref{eq:Force}, can be written as:

\begin{widetext}
\begin{equation}\label{eq:ForceRen}
\begin{aligned}
    f_{\textit{abinitio}} - f_{ren.}^{2nd} =  -\sum\limits_{{N_2 \nu},{N_3 \pi}} \sum\limits_{\beta,\gamma} \frac{1}{2!} \Psi_3^{\alpha \beta \gamma}({0 \mu}, {N_2 \nu}, {N_3 \pi}) u_{\beta}(N_2 \nu) u_{\gamma}(N_3 \pi)  - \sum\limits_{{N_2 \nu},{N_3 \pi},{N_4 \rho}} \sum\limits_{\beta,\gamma,\delta} \frac{1}{3!} \Psi_4^{\alpha \beta \gamma \delta} ({0 \mu}, {N_2 \nu}, {N_3 \pi}, {N_4 \rho})   \\ \times u_{\beta}(N_2 \nu) 
    u_{\gamma}(N_3 \pi) u_{\delta}(N_4 \rho)
\end{aligned} 
\end{equation}
By fitting Eq. \ref{eq:ForceRen} with the displacement dataset already obtained from TSS (or AIMD), we can determine the renormalized third- ($\Psi_3^{\alpha \beta \gamma}({0 \mu}, {N_2 \nu}, {N_3 \pi})$) and fourth-order ($\Psi_4^{\alpha \beta \gamma \delta} ({0 \mu}, {N_2 \nu}, {N_3 \pi}, {N_4 \rho})$) interatomic force constants.
\end{widetext}

\subsection{\label{sec:FreeEnergy}Free Energy}
 The expression of Helmholtz free energy up to fourth order anharmonicity is given by\cite{Wallace}: 
\begin{equation}\label{eq:FreeEnergy}
\begin{aligned}
    F = U_0 + F_H + F_A,
\end{aligned} 
\end{equation}
where harmonic free energy, 
\begin{equation}\label{eq:HarmonicFreeEnergy}
\begin{aligned}
    F_H = \sum\limits_{q, s} \left[ \frac{1}{2} \hbar \omega_{qs} + k_B T ln\left( 1 - e ^{-\frac{\hbar \omega_{qs}}{k_B T}} \right)\right],
\end{aligned} 
\end{equation}
and the anharmonic contribution to free energy:
\begin{widetext}
\begin{equation}\label{eq:AnHarmonicFreeEnergy}
\begin{aligned}
    F_A = 12 \sum\limits_{qs, q^{'}s^{'}} \phi(qs, -qs, q^{'}s^{'}, -q^{'}s^{'}) \left( n_{qs} + \frac{1}{2}\right) \left( n_{q^{'}s^{'}} + \frac{1}{2}\right) - \\ \frac{18}{\hbar} \sum\limits_{qs, q^{'}s^{'}, q^{''}s^{''}} \Bigg[ |\phi(qs, q^{'}s^{'}, q^{''}s^{''})|^2 \left\{ \frac{ n_{qs} n_{q^{'}s^{'}} + n_{qs} + \frac{1}{3} } {(\omega_{qs} + \omega_{q^{'}s^{'}} + \omega_{q^{''}s^{''}})_p}  + \frac{ 2n_{qs} n_{q^{''}s^{''}} - n_{qs} n_{q^{'}s^{'}} + n_{q^{''}s^{''}} } {(\omega_{qs} + \omega_{q^{'}s^{'}} - \omega_{q^{''}s^{''}})_p} \right\} + \\
    2 \phi(qs, -qs, q^{''}s^{''}) \phi(q^{'}s^{'}, -q^{'}s^{'}, -q^{''}s^{''})  \frac{ n_{qs} n_{q^{'}s^{'}} + n_{qs} + \frac{1}{4} } {(\omega_{q^{''}s^{''}})_p} \Bigg].
\end{aligned} 
\end{equation}
Here $\phi(qs, q^{'}s^{'}, q^{''}s^{''})$ and $\phi(qs, q^{'}s^{'}, q^{''}s^{''}, q^{'''}s^{'''})$ are scattering matrix elements related to third and fourth-order IFCs:
\begin{equation}\label{eq:ScatterProb3}
\begin{aligned}
    \phi(qs, q^{'}s^{'}, q^{''}s^{''}) = \frac{1}{3!} \left( \frac{\hbar}{2}\right)^{3/2} \frac{1}{\sqrt{N_0}}\frac{1}{\sqrt{\omega_{qs} \omega_{q^{'}s^{'}}\omega_{q^{''}s^{''}}}} \sum\limits_{\mu, {N_2 \nu},{N_3 \pi}} \sum\limits_{\alpha,\beta,\gamma} \Phi_3^{\alpha \beta \gamma}({0 \mu}, {N_2 \nu}, {N_3 \pi}) e^{i \Vec{q^{'}}\cdot\Vec{R(N_2)}} e^{i \Vec{q^{''}}\cdot\Vec{R(N_3)}} \times \\
    \frac{\epsilon_\alpha (qs, \mu) \epsilon_\beta (q^{'}s^{'}, \nu) \epsilon_\gamma (q^{''}s^{''}, \pi)}{\sqrt{M_\mu M_\nu M_\pi}},
\end{aligned} 
\end{equation}
and,
\begin{equation}\label{eq:ScatterProb4}
\begin{aligned}
    \phi(qs, q^{'}s^{'}, q^{''}s^{''}, q^{'''}s^{'''}) = \frac{1}{4!} \left( \frac{\hbar}{2}\right)^2 \frac{1}{N_0}\frac{1}{\sqrt{\omega_{qs} \omega_{q^{'}s^{'}}\omega_{q^{''}s^{''}} \omega_{q^{'''}s^{'''}} }} \sum\limits_{\mu, {N_2 \nu},{N_3 \pi}, {N_4 \rho}} \sum\limits_{\alpha,\beta,\gamma, \delta} \Phi_4^{\alpha \beta \gamma \delta}({0 \mu}, {N_2 \nu}, {N_3 \pi}, {N_4 \rho}) \times \\
    e^{i \Vec{q^{'}}\cdot\Vec{R(N_2)}} e^{i \Vec{q^{''}}\cdot\Vec{R(N_3)}} e^{i \Vec{q^{'''}}\cdot\Vec{R(N_4)}} \times \frac{\epsilon_\alpha (qs, \mu) \epsilon_\beta (q^{'}s^{'}, \nu) \epsilon_{\gamma}(q^{''}s^{''}, \pi)  \epsilon_\delta (q^{'''}s^{'''}, \rho)}{\sqrt{M_\mu M_\nu M_\pi M_\rho}}.
\end{aligned} 
\end{equation}
\end{widetext}
The given expressions for scattering matrix elements incorporate quasi-momentum conservation, where the relations $q + q^{'} + q^{''} = G$ and $q + q^{'} + q^{''} + q^{'''} = G$ hold, with $G$ representing the reciprocal lattice vector of the crystal. 
\subsection{\label{sec:ThermalConductivity}Thermal Conductivity}
 
The non-equilibrium distribution function obtained from the solution of the linearized  Peierls-Boltzmann transport equation (PBTE) is used to calculate thermal conductivity ($\kappa$).
The sources of the scattering term in PBTE are currently limited to three-phonon (3-ph), four-phonon (4-ph), and isotopic impurity\cite{Isotope_sctr} scattering in our implementation. 
The three-phonon scattering rate, as given by Fermi golden rule: 
\begin{widetext}
\begin{equation}\label{eq:3-ph}
 \begin{aligned}
     \frac{1}{\tau_{3-ph}(qs)} = \frac{2\pi}{\hbar^2}\sum\limits_{q_1s_1, q_2s_2} \Bigg[ \lvert \phi(qs, q_1s_1, -q_2s_2) \rvert ^2 (n_{q_1s_1} - n_{q_2s_2}) \delta (\omega_{qs} + \omega_{q_1s_1} - \omega_{q_2s_2}) + \frac{1}{2} \lvert \phi(qs, -q_1s_1, -q_2s_2) \rvert ^2 \times \\  (1 + n_{q_1s_1} + n_{q_2s_2}) \delta (\omega_{qs} - \omega_{q_1s_1} - \omega_{q_2s_2}) \Bigg], 
 \end{aligned}   
\end{equation}
the four-phonon scattering rate:
\begin{equation}\label{eq:4-ph}
 \begin{aligned}
 \frac{1}{\tau_{4-ph}(qs)} = \frac{2\pi}{\hbar^2}\sum\limits_{q_1s_1, q_2s_2, q_3s_3} \Bigg[ \frac{1}{6} \lvert \phi(qs, -q_1s_1, -q_2s_2, -q_3s_3) \rvert ^2 \frac{n_{q_1s_1} n_{q_2s_2} n_{q_3s_3}} {n_{qs}} \delta (\omega_{qs} - \omega_{q_1s_1} - \omega_{q_2s_2} - \omega_{q_3s_3}) + \\ 
 \frac{1}{2} \lvert \phi(qs, q_1s_1, -q_2s_2, -q_3s_3) \rvert ^2 \frac{(1+n_{q_1s_1}) n_{q_2s_2} n_{q_3s_3}} {n_{qs}} \delta (\omega_{qs} + \omega_{q_1s_1} - \omega_{q_2s_2} - \omega_{q_3s_3}) + \\
 \frac{1}{2} \lvert \phi(qs, q_1s_1, q_2s_2, -q_3s_3) \rvert ^2 \frac{(1+n_{q_1s_1}) (1+n_{q_2s_2}) n_{q_3s_3}} {n_{qs}} \delta (\omega_{qs} + \omega_{q_1s_1} + \omega_{q_2s_2} - \omega_{q_3s_3}) \Bigg],
 \end{aligned}   
\end{equation}
and the phonon-isotopic impurity scattering rate:
\begin{equation}\label{eq:ph-iso}
 \begin{aligned}
 \frac{1}{\tau_{ph-iso}(qs)} = \frac{\pi \omega_{qs}^2}{2N_0}\sum\limits_{q_1s_1} \sum\limits_{\mu} g_2(\mu) \: \lvert \epsilon (qs, \mu)^* \cdot  \epsilon (q_1s_1, \mu) \rvert ^2 \delta (\omega_{qs} - \omega_{q_1s_1} ),
 \end{aligned}   
\end{equation}
\end{widetext}
where $g_2(\mu) = \sum\limits_{i} f_{\mu i} (1-M_{\mu i}/\bar{M_\mu})^2$, with $M_{\mu i}$ and $f_{\mu i}$ being the mass and fractional concentration of the ith isotope of $\mu$th basis atom and $\bar{M_\mu} = \sum\limits_{i} f_{\mu i} M_{\mu i}$ is the average mass. 
The total scattering rate, as given by Matthiessen’s rule: 
\begin{equation}\label{eq:Sctr-rate}
 \begin{aligned}
     & \frac{1}{\tau(qs)} = \frac{1}{\tau_{3-ph}(qs)} + \frac{1}{\tau_{4-ph}(qs)} + \frac{1}{\tau_{ph-iso}(qs)} .
 \end{aligned}   
\end{equation}
The expression for thermal conductivity tensor ($\boldsymbol{\kappa}$) in relaxation time approximation (RTA) is given by: 
\begin{equation}\label{eq:kappa-RTA}
 \begin{aligned}
     & \boldsymbol{\kappa ^{RTA}} = \frac{1}{N_0 V_0} \sum\limits_{q, s} \frac{\partial n_{qs}}{\partial T} \: \hbar \omega_{qs} \: \tau(qs) \: \vec{v}_{qs} \otimes \vec{v}_{qs}.
 \end{aligned}   
\end{equation}
The thermal conductivity tensor from the iterative solution of PBTE is represented by the following expression:
\begin{equation}\label{eq:kappa}
 \begin{aligned}
     & \boldsymbol{\kappa} = \frac{k_B T^2}{N_0 V_0} \sum\limits_{q, s} \frac{\partial n_{qs}}{\partial T} \vec{v}_{qs} \otimes \vec{\Delta}_{qs}
 \end{aligned}   
\end{equation}
Here $N_0$ is the number of q-points in the BZ sampled, $V_0$ is the unit cell volume, and $\vec{v}_{qs} = \frac{1}{2 \omega_{qs}} \langle \epsilon(qs) \vert \frac{\partial D}{\partial \vec{q}} \vert \epsilon(qs) \rangle$ is phonon group velocity with $D$ being dynamical matrix (Eq. \ref{eq:DynMat}). 
The self-consistent equation governing the vector function ($\vec{\Delta}_{qs}$) is interconnected with the phonon scattering rate and 3-ph, 4-ph, and phonon-isotope scattering probabilities.
For further details regarding these terms, readers can consult the following  references\cite{Ziman, BTE_lin_iter, Lindsay_2008, Ravichandran_2018}. 
The novel aspects of our theoretical framework  are as follows: (1) The iterative solution of PBTE\cite{itr_BTE}, which differentiates the momentum-conserving normal processes as non-resistive. 
(2) The calculation of phonon group velocity involves the exact analytical differentiation of the long-range Ewald term\cite{EwaldNonAnalytic2} for polar materials. This distinguishes our implementation from the others, which typically differentiate direction-dependent non-analytic terms for determining phonon group velocity in polar materials.
(3) In the treatment of energy conservation within the Brillouin zone (BZ) sum, the tetrahedron method\cite{Tetrahedron} is employed. The tetrahedron method offers several advantages over the Gaussian or Lorentzian approximation of the Dirac-delta function. One notable advantage is that it does not rely on any adjustable parameters and exhibits slightly improved convergence with the number of q-points.
In this study, the convergence of BZ integration for thermal conductivity is achieved using a q-mesh grid of $17\times17\times17$ for the 3-phonon scattering processes.
The phonon linewidth ($\Gamma (qs)$) discussed in this paper is defined as:
\begin{equation}\label{eq:LW}
 \begin{aligned}
     & \Gamma (qs) = \frac{1}{2 \: \tau (qs)}.
 \end{aligned}   
\end{equation}
For all the {\em ab initio} calculations, we utilized SIESTA\cite{SIESTA_2002}, a code based on atomic-orbital basis. 
To expand the wave functions, we employ the polarized double-zeta basis set. 
Norm-conserving pseudopotentials are utilized, and exchange-correlation treatment follows the GGA-PBE\cite{GGA-PBE} approach.

\begin{figure*}[hbt!]
    \hspace{-3.5ex}
    \includegraphics[width=18.4cm]{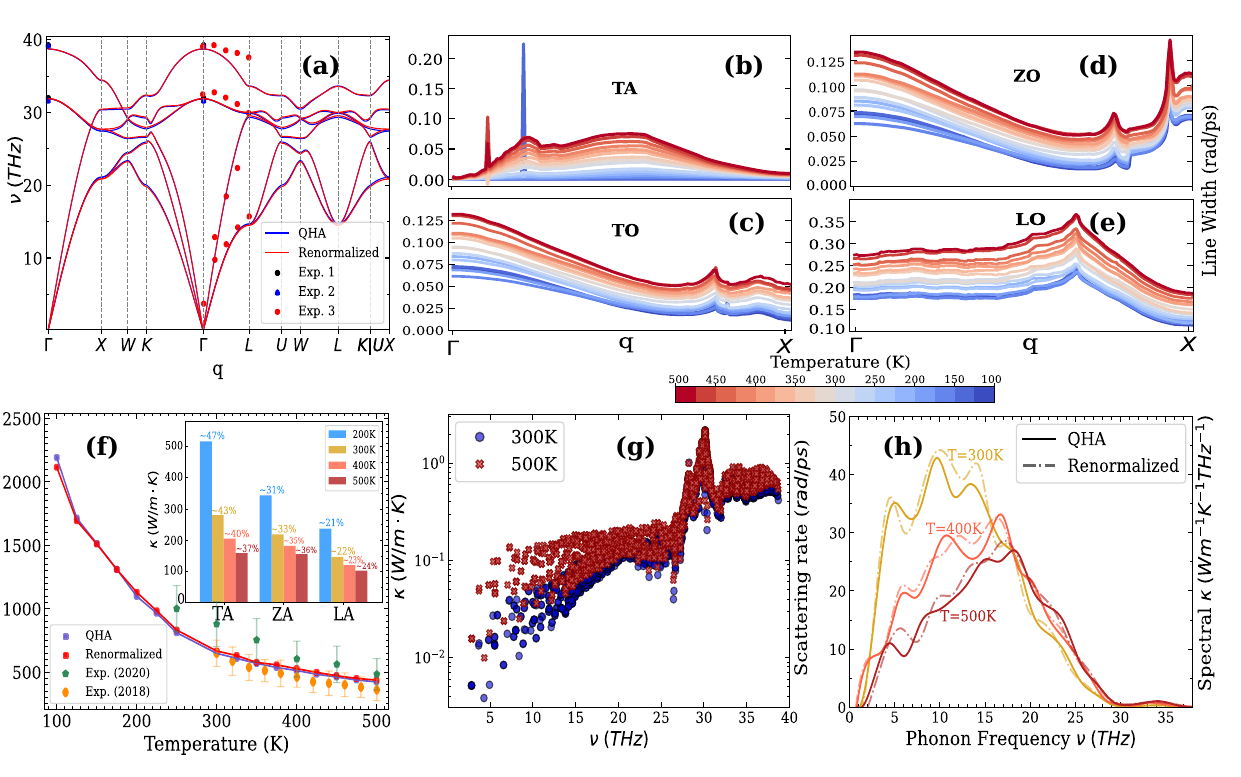}
    \caption{(a) Phonon dispersion of cBN along the high symmetry path calculated from QHA and renormalized IFCs at 300K. Observe that there is no significant difference between QHA and renormalization. The experimental data are extracted from Exp.1\cite{BN_phonon_disp_Exp1}, Exp.2\cite{BN_phonon_disp_Exp2}, and Exp.3\cite{BN_phonon_disp_Exp3}. 
(b) - (e) Variation in phonon line width along the $\Gamma \to X$ high-symmetry path for one acoustic branch (TA) and three optic branches (TO, ZO, and LO) in cBN across temperatures ranging from 100K to 500K. 
(f) Thermal conductivity of cBN calculated across the temperature range of 100K to 500K. Renormalization results in approximately a 2\% to 3\% difference in the $\kappa$ value compared to the QHA. Experimental data are obtained from Exp. (2020)\cite{Exp_kappa_2020} and Exp. (2018)\cite{Exp_kappa_2018}. Inset plot: Contribution from the three acoustic phonon branches to the thermal conductivity at different temperatures.
(g) Phonon scattering rates, considering 3-phonon and phonon-isotope scattering, calculated for cBN at 300K and 500K using renormalized IFCs. The elevated scattering rates at higher temperatures contribute to a reduction in thermal conductivity.
(h) Spectrally resolved thermal conductivity of cBN at various temperatures, providing insights into the scattering rate plot. Notably, the renormalized spectral $\kappa$ exhibits a slight increase across all temperatures.}
    \label{fig:Fig_1}
\end{figure*}

\section{\label{sec:Results} RESULTS}

\subsection{\label{sec:cBN_3C-SiC} High $\kappa$ Material: cBN and 3C-SiC}
\begin{figure*}[hbt!]
    \hspace{-3.5ex}
    \includegraphics[width=18.4cm]{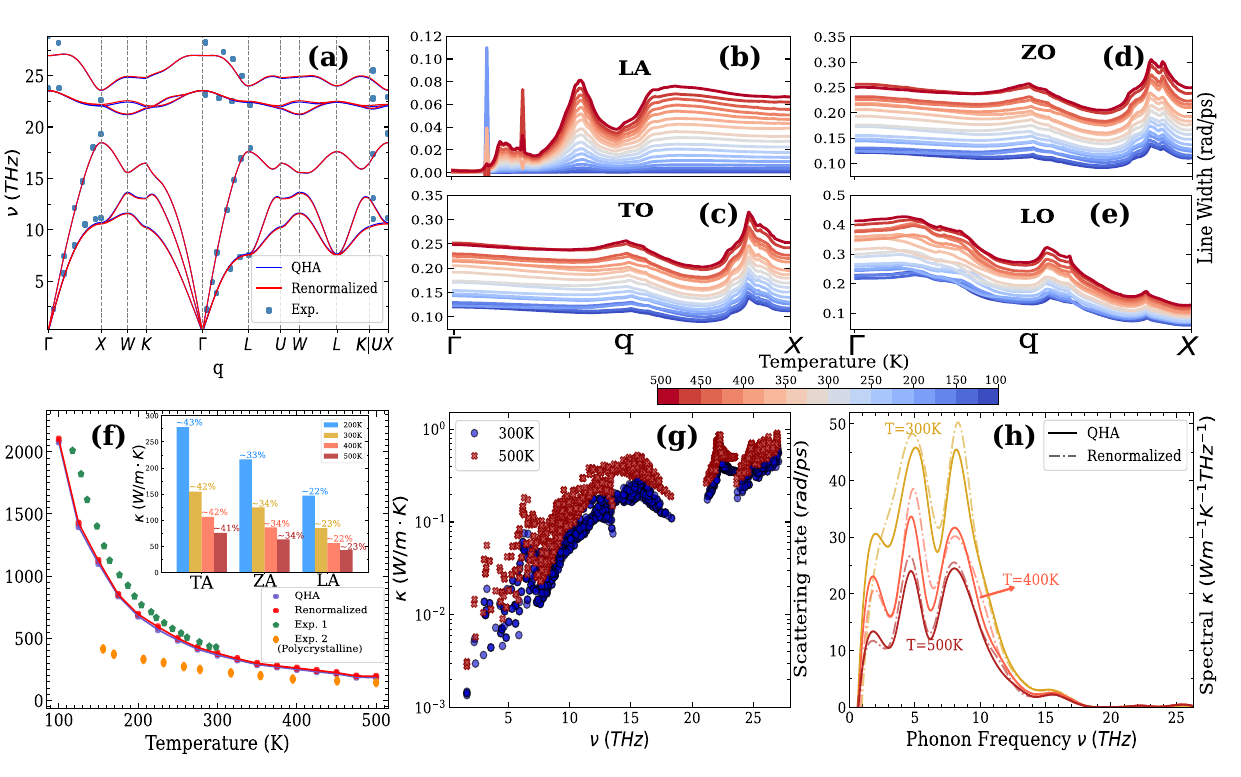}
    \caption{(a) Phonon dispersion of 3C-SiC along the high symmetry path calculated from QHA and renormalized IFCs at 300K. Observe that there is no significant difference between QHA and renormalization. The experimental data are extracted from Ref.\cite{SiC_phonon_disp} 
(b) - (e) Variation in phonon linewidth along the $\Gamma \to X$ high-symmetry path for one acoustic branch (LA) and three optic branches (TO, ZO, and LO) in 3C-SiC across temperatures ranging from 100K to 500K. 
(f) Thermal conductivity of 3C-SiC calculated across the temperature range of 100K to 500K. Renormalization results in a marginally higher $\kappa$ value compared to the QHA. Experimental data are obtained from Exp.1\cite{kappa_singleCrystallineSiC} and Exp.2\cite{kappa_polyCrystallineSiC}. Note that Exp.2\cite{kappa_polyCrystallineSiC} corresponds to a polycrystalline sample, which tends to match better with our calculation at higher temperatures. Inset plot: Contribution from the three acoustic phonon branches to the thermal conductivity at different temperatures.
(g) Phonon scattering rates, considering 3-phonon + phonon-isotope scattering, calculated for 3C-SiC at 300K and 500K using renormalized IFCs. The elevated scattering rates at higher temperatures contribute to a reduction in thermal conductivity.
(h) Spectrally resolved thermal conductivity of 3C-SiC at various temperatures. Notably, the renormalized spectral $\kappa$ exhibits a slight increase across all temperatures.}
    \label{fig:Fig_2}
\end{figure*}

We selected cubic-Boron Nitride (cBN) and cubic-Silicon Carbide (3C-SiC) as materials characterized by low anharmonicity in their interatomic potential. 
High thermal conductivity in these materials is attributed to the combination of weak anharmonicity and the light atomic mass of their constituent atoms. 
Figure \ref{fig:Fig_1}(a) and Figure \ref{fig:Fig_2}(a) display phonon dispersion for cBN and 3C-SiC, respectively, presenting results from both the QHA and renormalized IFCs at 300K.
Note that in QHA, the temperature effect on the lattice constant is taken into account, and the displacement-force dataset (Eq. \ref{eq:Force}) is fitted up to the fourth order.
There is little to no noticeable difference between the phonon frequencies obtained through the QHA and the renormalized approach.
This signifies the existence of weak higher-order anharmonic terms in these materials, which appear in the renormalization process. 
The figures also include the experimental phonon frequency for cBN\cite{BN_phonon_disp_Exp1, BN_phonon_disp_Exp2, BN_phonon_disp_Exp3} and 3C-SiC\cite{SiC_phonon_disp}, demonstrating a close agreement with our calculations.
Figure \ref{fig:Fig_1}(b-e) and Figure \ref{fig:Fig_2}(b-e) illustrate the temperature-dependent phonon linewidth for one acoustic branch and optic branches along the high-symmetry $\Gamma \to X$ path of cBN and 3C-SiC, respectively.
As depicted in the figures, the phonon linewidth exhibits a monotonic increase with temperature, resulting in reduced thermal transport as the temperature rises.
Observe that the phonon linewidth for the acoustic branch is significantly smaller compared to that of the optic branches, suggesting lower scattering for the acoustic phonons.
This difference contributes to the high thermal conductivity of acoustic phonons, given the inverse relationship between thermal conductivity and linewidth.
More than 90\% of the total $\kappa$ value for both materials is contributed by three acoustic modes, as depicted in the inset plot of Figure \ref{fig:Fig_1}(f) and Figure \ref{fig:Fig_2}(f).
Also, observe that certain peak positions occur within the linewidth curve along the high-symmetry path. These peaks emerge as both momentum conservation (Eq. \ref{eq:ScatterProb3}) and energy conservation (Eq. \ref{eq:3-ph}) are met to a greater extent, leading to heightened scattering for these phonon modes ($qs$). Consequently, the linewidth of these phonon modes becomes more pronounced, manifesting as peaks within the linewidth curve.
The temperature-dependent thermal conductivity of cBN and 3C-SiC is presented in Figure \ref{fig:Fig_1}(f) and Figure \ref{fig:Fig_2}(f) across the 100K to 500K range. 
Note that for these materials characterized by high thermal conductivity, renormalization of the IFCs has minimal to no impact on the $\kappa$ values.
For instance, cBN exhibits a thermal conductivity of 648.31 W/m.K at 300K, according to the QHA. 
Meanwhile, the renormalization of IFCs predicts $\kappa$ at 666.97 W/m.K, representing a roughly 3\% increase compared to the QHA-derived value. 
The experimental determination of the $\kappa$ of cBN is 648$\pm$52 W/m.K as extracted from Ref.\cite{Exp_kappa_2018} and 850 $\pm$ 90 W/m.K as extracted from Ref.\cite{Exp_kappa_2020} 
Note that in our calculation of phonon-isotope scattering, we considered the natural abundance of the isotopes of the constituent atoms, and the experimental values extracted also correspond to the natural isotopic concentration in the sample.
For 3C-SiC, our calculation using the QHA yields a thermal conductivity value of 364.25 W/m.K, while the renormalization predicts a slightly elevated $\kappa$ of 378.68 W/m.K at 300K.
The experimental $\kappa$ of single-crystalline 3C-SiC as extracted from the Ref.\cite{kappa_singleCrystallineSiC} is 437 W/m.K.
Figure \ref{fig:Fig_2}(f) also includes the thermal conductivity of polycrystalline 3C-SiC, as extracted from the Ref.\cite{kappa_polyCrystallineSiC}, which tends to match our calculation at higher temperatures. 
The bar plots inset in Figure \ref{fig:Fig_1}(f) and Figure \ref{fig:Fig_2}(f) demonstrate the contribution of the three acoustic phonon branches to thermal conductivity at various temperatures. 
With increasing temperature, higher-frequency phonon branches and optic branches tend to contribute more to thermal conduction.
Figure \ref{fig:Fig_1}(g) and Figure \ref{fig:Fig_2}(g) illustrate the three-phonon + phonon-isotope scattering rates at temperatures of 300K and 500K for cBN and 3C-SiC, respectively.
The elevation in scattering rate as temperature increases accounts for the reduction in thermal conductivity.
To translate the scattering rate plot into thermal conductivity, we computed spectral thermal conductivity, as shown in Figure \ref{fig:Fig_1}(h) and Figure \ref{fig:Fig_2}(h). 
Observe that the spectral thermal conductivity tends to decrease as temperature rises, resulting in a reduction in the overall thermal conductivity.
Additionally, note that the spectral thermal conductivity for renormalized IFCs is slightly higher than that of the QHA for all the temperatures. This accounts for the marginal increase in thermal conductivity observed with the renormalized approach in our calculations. 

\subsection{\label{sec:NaCl_AgI} Low $\kappa$ Material: NaCl and AgI}
\begin{figure*}[hbt!]
	  \centering
	\includegraphics[width=18.5cm, height=10.5cm]{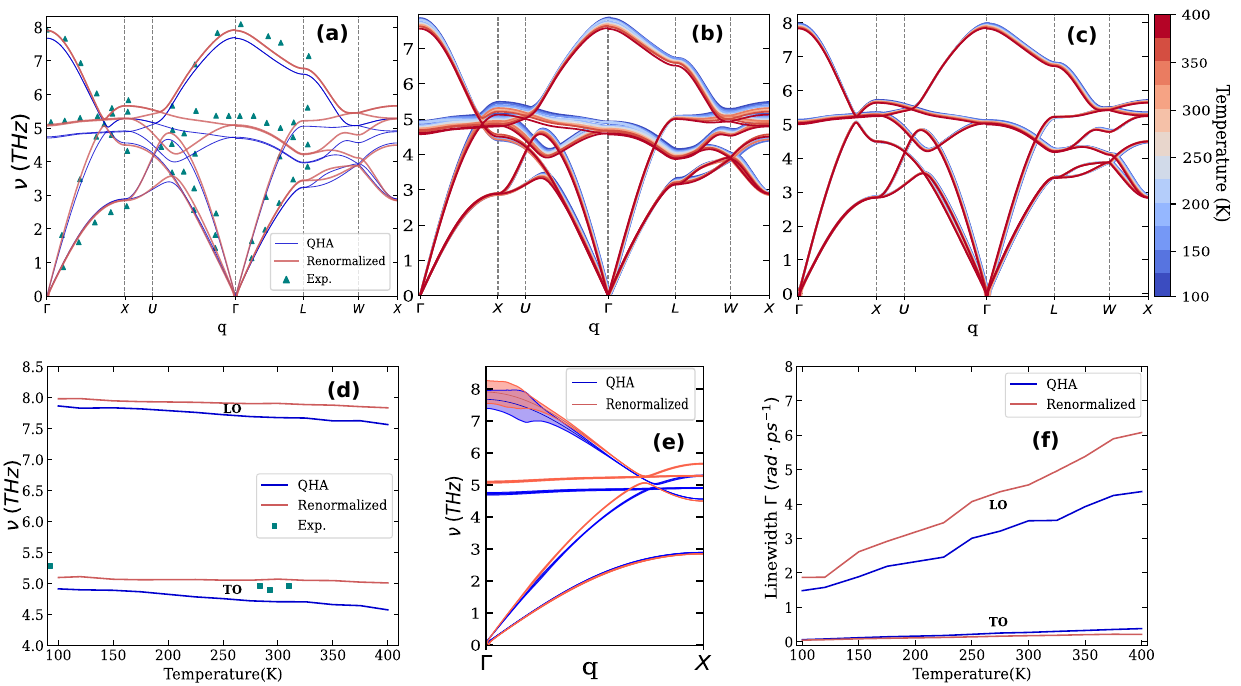}
    \caption{(a) Phonon dispersion of NaCl along the high-symmetry path at 300K. The renormalization leads to an upward shift in the frequency of the optical branches, aligning with the experimental data.\cite{Raunio_PhononDisp} (b) Temperature dependence of the bare phonon dispersion within the 100K-400K range. Observe the pronounced softening of optical branches with increasing temperature. (c) The renormalization leads to weaker temperature dependence phonon dispersion consistent with the experimental data. (d) Temperature dependence of the phonon frequency at the $\Gamma$ point for LO and TO branches. Notice the reduced extent of softening with increasing temperature due to renormalization, closely matching the experimental data.\cite{Linewidth} (e) Distribution of phonon linewidth along the $\Gamma$-X path at 300K. The vertical spread of the shaded region surrounding each phonon branch determines the linewidth.  (f) Temperature dependence of the phonon linewidth at the $\Gamma$ point for LO and TO branches.}
    \label{fig:Fig_3}
\end{figure*}

To demonstrate the practicality of our numerical framework for materials with low thermal conductivity and significant anharmonicity, we selected sodium chloride (NaCl) and silver iodide ($\gamma$-AgI, space group - $F\Bar{4}3m$) as representative materials.
Being a member of the alkali halides family, NaCl exhibits high anharmonicity in its inter-atomic potential due to the weakly bonded, highly ionic nature of its constituent atoms.
Furthermore, it displays remarkably low thermal conductivity, making it an ideal candidate for studying the anharmonicity within our current framework. 
Figure \ref{fig:Fig_3}(a) presents the phonon dispersion of NaCl crystal at 300K, calculated from QHA and through the renormalization of the phonon quasiparticle as described in the method section. 
Our findings indicate that the renormalization process significantly increases the phonon frequency of the optical modes of NaCl. Such behavior is not observed in cBN and 3C-SiC. 
The acoustic branches remain largely unaffected after renormalization.
The experimentally measured data\cite{Raunio_PhononDisp} through the inelastic neutron scattering demonstrates a close agreement with the renormalized phonon frequency. 
This emphasizes the significance of treating phonons as quasiparticles through the renormalization process, rather than considering them solely as normal modes of lattice vibrations, especially for highly anharmonic materials.
Figure \ref{fig:Fig_3}(b) to Figure \ref{fig:Fig_3}(d) show the dependence of phonon dispersion with temperature from 100K to 400K, both for bare phonon within QHA and renormalized phonon. 
The outcome reveals that, in comparison to the renormalized phonon, the bare phonon experiences a higher degree of softening in the optical phonon branches as the temperature increases. 
In Figure \ref{fig:Fig_3}(d), the plot illustrates the variation of phonon frequency at the $\Gamma$-point of the Brillouin zone (BZ) with temperature for both the transverse optical (TO) and longitudinal optical (LO) branches.
In contrast to QHA, the renormalized phonon demonstrates a minimal temperature dependence while remaining consistent with the experimental data.
Figure \ref{fig:Fig_3}(e) illustrates the distribution of phonon linewidth or inverse phonon lifetime along the $\Gamma$-X high-symmetry path in the BZ due to the three-phonon scattering process at 300K. 
The quantitative value of the linewidth can be obtained from the vertical extent of the highlighted area around the phonon branch, with the unit being the same as the phonon frequency (THz).
In Figure \ref{fig:Fig_3}(f), it can be observed that for the temperature range of 100K-400K, the linewidth of the LO branch at the $\Gamma$-point is greater in the case of the renormalized phonon.
Conversely, the renormalized linewidth of the TO branch consistently remains lower compared to that of the bare phonon.
It should be noted that, in our approach, anharmonic third-order IFCs are also subject to renormalization, according to Eq. 5.
The increased linewidth of the LO branch indicates a higher rate of 3-phonon scattering, resulting in a reduced contribution to thermal conduction.
\begin{figure*}[hbt!]
    \includegraphics[width=18.3cm]{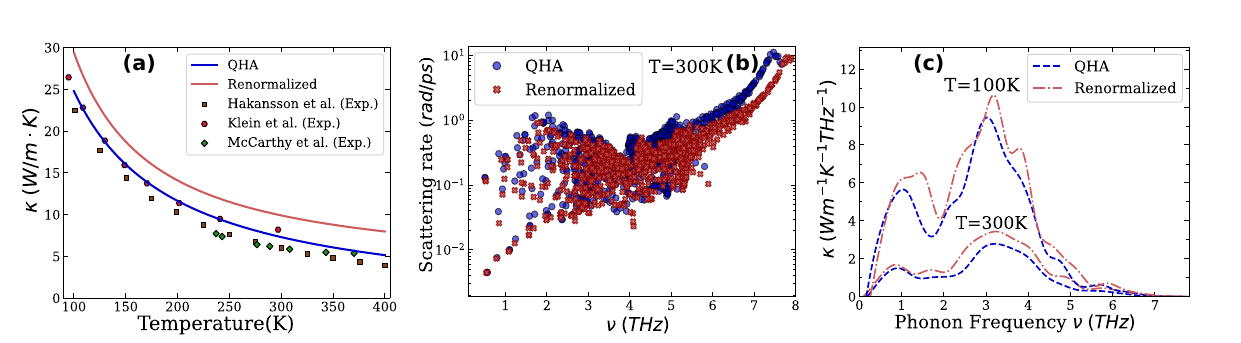}
    \caption{(a) Thermal conductivity of NaCl calculated using the QHA and renormalization. Notably, renormalization consistently overestimates the thermal conductivity across all temperatures. Experimental values are extracted from Ref.\cite{HAKANSSON_Exp_Pressure, Klein_Exp, McCarthy_Exp} (b) Three-phonon scattering rate calculated from bare and renormalized IFCs at 300K. (c) Spectral distribution of thermal conductivity at 100K and 300K calculated through QHA and renormalization. }
    \label{fig:Fig_4}
\end{figure*}
Figure \ref{fig:Fig_4}(a) presents the thermal conductivity ($\kappa$) in the 100K to 400K temperature range calculated from bare and renormalized IFCs. 
The thermal conductivity calculated from the three-phonon scattering process of the renormalized phonon consistently overestimates the values obtained from QHA and experimental data.
For example, the renormalized $\kappa$ at 100K is 28.78 W/mK, exhibiting an approximately 17\% increase compared to the unrenormalized $\kappa$ of 24.62 W/mK. 
At 300K, the renormalization results in an overprediction of about 25\%. 
\begin{figure*}[hbt!]
    \hspace{-3.5ex}
    \includegraphics[width=18.4cm]{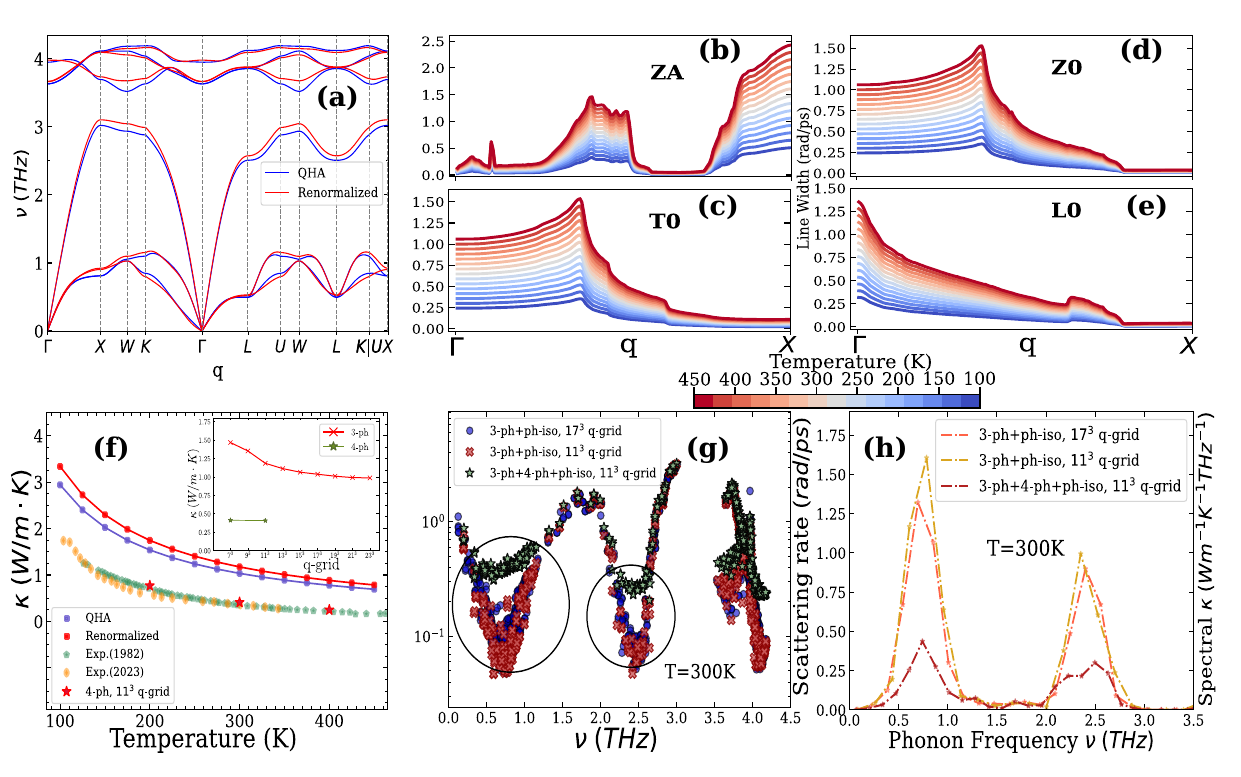}
    \caption{(a) Phonon dispersion of AgI along the high symmetry path calculated from QHA and renormalized IFCs at 300K. 
(b) - (e) Variation in phonon linewidth for one acoustic branch (ZA) and three optic branches (TO, ZO, and LO) in AgI across temperatures ranging from 100K to 450K. 
(f) Thermal conductivity of AgI calculated over the temperature range of 100K to 450K using 3-phonon scattering. Results from four-phonon scattering processes are presented for the $11^3$ q-grid for 200 K, 300 K, and 400 K temperatures. Experimental data are sourced from Exp. (1982)\cite{AgI_Exp_1982} and Exp. (2023)\cite{AgI_2023_kappa_1}. It is noteworthy that both the QHA and renormalized values tend to over-predict the measured $\kappa$ values in three-phonon, whereas the four-phonon prediction closely aligns with the experimental data. Inset plot: Convergence of $\kappa$ with q-grid in BZ for three-phonon and four-phonon processes.
(g) Phonon scattering rates, taking into account 3-phonon + phonon-isotope scattering at $11^3$ and $17^3$ q-grid, and 3-phonon + 4-phonon + phonon-isotope scattering processes at $11^3$ q-grid. The highlighted regions indicate a significant increase in the scattering rate, approximately an order of magnitude higher for four-phonon processes compared to three-phonon processes. Notably, the two dips in the scattering rate in the highlighted region correspond to the two peaks in the spectral $\kappa$.
(h) Spectrally resolved thermal conductivity of AgI for three-phonon and four-phonon scattering, illustrating the notable decrease in $\kappa$ when fourth-order processes are considered.
}
    \label{fig:Fig_5}
\end{figure*}
Importantly, our implementation of the tetrahedron method for energy conservation delta function does not involve any adjustable parameters for the thermal conductivity calculation, distinguishing it from the Gaussian or Lorentzian approximations. 
On the other hand, the thermal conductivity obtained from the bare IFCs exhibits a close agreement with the experimental data, even in the scenario involving minimal three-phonon scattering processes.
At a temperature of 300K, the $\kappa$ derived from the QHA is in close proximity to the experimental results, with a deviation of less than 7\% from the findings in Ref\cite{Klein_Exp} and less than 16\% from Ref\cite{HAKANSSON_Exp_Pressure}. 
To interpret this anomalous behavior of renormalized phonon, we calculated the three-phonon scattering rate from renormalized and bare IFCs, as presented in Figure \ref{fig:Fig_4}(b). 
As illustrated in the figure, the scattering rate derived from the bare IFCs surpasses the scattering rate of the renormalized IFCs.
Furthermore, the bare phonon exhibits a greater extent of temperature-induced softening, as depicted in Figure \ref{fig:Fig_3}(a-d).
By combining these two factors, the thermal conductivity predicted from the unrenormalized IFCs consistently remains lower than that from the renormalized IFCs, since $\kappa$ is directly proportional to the phonon frequency and inversely proportional to the scattering rate.
This can be observed in Figure \ref{fig:Fig_4}(c), which illustrates the spectral distribution of $\kappa$ at 100K and 300 K.
At both temperatures, the renormalized spectral $\kappa$ consistently surpasses the values obtained from the unrenormalized spectral $\kappa$ across all phonon frequency ranges, reinforcing the higher thermal conductivity exhibited by the renormalized phonons.
The peculiar characteristics of the renormalized phonon can be addressed by incorporating higher-order phonon scattering processes, as elaborated in the Ref.\cite{Ravichandran_2018}, especially for highly anharmonic materials like NaCl.
The inclusion of four-phonon scattering processes provides a means to address this issue.

Figure \ref{fig:Fig_5} presents the thermal transport-related physical properties of AgI.
Research interest has recently grown in the thermal transport of materials exhibiting ultralow thermal conductivity\cite{Kanishka2022, Kanishka2023}, primarily due to their potential applications in thermoelectric devices. Among these materials, AgI stands out as a promising candidate.
Understanding the ultralow thermal transport in AgI has been enhanced by recent studies\cite{AgI_2023_kappa_1, AgI_2023_kappa_2} highlighting the significance of higher-order phonon scattering processes. 
Figure \ref{fig:Fig_5}(a) presents the phonon dispersion of AgI along the high symmetry path of BZ at 300K calculated from both the unrenormalized and renormalized IFCs. 
Note that the phonon frequencies experience a slight stiffening across both the acoustic and optical branches when considering renormalized IFCs. 
The increase in phonon frequency in AgI resulting from renormalization is of a smaller magnitude than that observed in NaCl.
Figure \ref{fig:Fig_5}(b-e) depicts temperature-dependent phonon linewidth for one acoustic branch and three optic branches along the high symmetry $\Gamma \to X$ path for the 100K to 450K temperature range. 
The monotonic increase of the linewidth with temperature implies higher phonon scattering as temperature rises.  
Figure \ref{fig:Fig_5}(f) presents the thermal conductivity of AgI for the temperature range 100K to 450K, calculated using both QHA and renormalized IFCs. 
The thermal conductivity calculations, relying solely on anharmonic 3-phonon and phonon-isotope scattering as the primary sources of scattering in PBTE, are represented by solid red and blue lines for QHA and renormalized IFCs, respectively.
The experimental values, as extracted from the Ref.\cite{ AgI_Exp_1982, AgI_2023_kappa_1}, are also presented in the figure. 
Observe that our calculated $\kappa$ values from three-phonon scattering processes consistently exceed the experimental values across all temperature ranges, and the process of renormalization leads to an additional overprediction compared to the QHA values.
For example, our calculated $\kappa$ of AgI at 300K is 1.03 W/m.K, whereas the experimental $\kappa$ is 0.36 W/m.K, as reported in Ref.\cite{AgI_2023_kappa_1} and 0.41 W/m.K as extracted from the Ref\cite{AgI_Exp_1982}. 
The renormalization predicts $\kappa$ as 1.17 W/m.K, which is about 13\% higher than the QHA value. 
The discrepancy between the calculated thermal conductivity and experimental observations highlights the significance of higher-order phonon scattering processes (specifically, 4-phonon interactions) for highly anharmonic materials such as AgI. 
To enhance the robustness of our numerical framework in handling materials with extreme anharmonicity, we incorporated the computation of four-phonon scattering processes for AgI.
Figure \ref{fig:Fig_5}(f) displays the computed thermal conductivity, taking into account four-phonon scattering, for temperatures of 200 K, 300 K, and 400 K.
Take note of the impressive alignment between the computed $\kappa$ and experimental values achieved simply by integrating the fourth-order scattering process.
Our calculation predicts the $\kappa$ of AgI to be 0.41 W/m.K at 300 K, closely resembling the experimental values of 0.41 W/m.K\cite{AgI_Exp_1982} and 0.36 W/m.K\cite{AgI_2023_kappa_1}.
The computational expense increases significantly by several orders of magnitude when incorporating four-phonon scattering.
To address this computational bottleneck, we mitigated it by leveraging symmetries in the calculation of fourth-order scattering matrix elements (Eq. \ref{eq:ScatterProb4}). 
In Eq. \ref{eq:ScatterProb4}, the first index ($qs$) is constrained to the irreducible Brillouin Zone (BZ), and the scattering matrix elements for permutations among the remaining three indices ($q^{'}s^{'}, q^{''}s^{''}, q^{'''}s^{'''}$) are computed only once, as they yield the same values.
We also utilized specific computer hardware acceleration techniques to enhance the speed of the computation.
To provide a comparison of computational costs, our three-phonon calculation for a $17\times17\times17$ ($17^3$) q-grid in the BZ required approximately 6 minutes, utilizing 240 CPU cores distributed across 5 compute nodes. In contrast, our four-phonon calculation for an $11\times11\times11$ ($11^3$) q-grid took 15 hours, employing 960 CPU cores distributed across 20 compute nodes.
The convergence of $\kappa$ with q-grid is depicted in the inset plot in Figure \ref{fig:Fig_5}(f).
While it requires a $17^3$ q-grid for three-phonon convergence, no significant difference in $\kappa$ values for four-phonon has been observed between $7^3$ and $11^3$ q-grid. The convergence of $\kappa$ at a lower q-grid for four-phonon processes has been well-established in previous studies\cite{FourPhonon, Ruan_4ph}.
In our calculation, we address energy conservation using the parameter-free tetrahedron method\cite{Tetrahedron}, which achieves convergence with a smaller q-grid compared to Gaussian or Lorentzian approximations.
In Figure \ref{fig:Fig_5}(g), the scattering rate is depicted for both three-phonon+phonon-isotope scattering processes and the additional contribution from fourth-order scattering mechanisms calculated through the Fermi golden rule.
Note that the scattering rates, encompassing four-phonon processes, are computed with an $11^3$ q-grid. To facilitate a direct comparison with three-phonon scattering, we have also included the three-phonon scattering rate for both $11^3$ q-grid and $17^3$ q-grid.
Note that the inclusion of four-phonon scattering results in an increased scattering rate compared to the conventional three-phonon scattering.
In the highlighted region of Figure \ref{fig:Fig_5}(g), the four-phonon scattering rate is approximately an order of magnitude greater than that of the three-phonon.
The rise in phonon scattering rates results in a decrease in the thermal transport coefficient of AgI during the four-phonon process, as depicted in Figure \ref{fig:Fig_5}(f).
For a more in-depth exploration of the impact of higher-order four-phonon scattering on thermal conductivity, we computed the spectral distribution of $\kappa$, as depicted in Figure \ref{fig:Fig_5}(h).
Observe that the two peaks in the spectral $\kappa$ align with the two dips in the highlighted region of the phonon scattering rate in Figure \ref{fig:Fig_5}(g).
The spectral $\kappa$ exhibits a notable reduction in the four-phonon process across all phonon frequency ranges, leading to an overall decline in the total thermal conductivity compared to the three-phonon scenario.

\begin{figure*}[hbt!]
    \hspace{-3.5ex}
    \includegraphics[width=18.4cm]{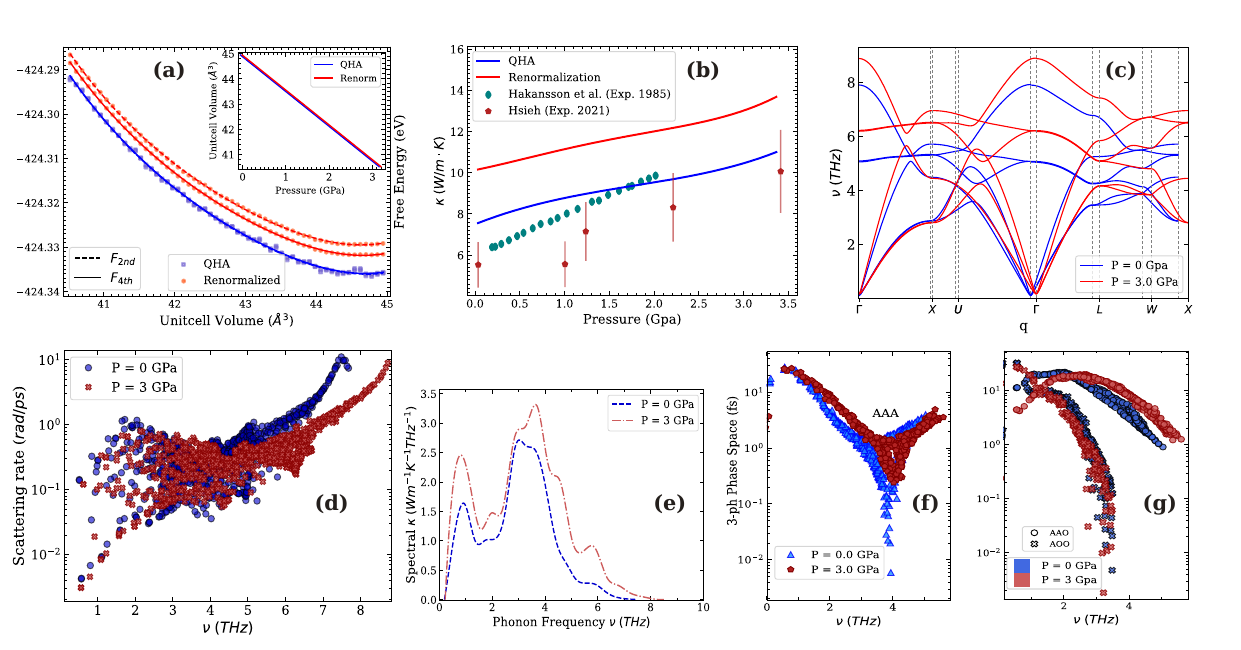}
    \caption{(a) Variation of harmonic and fourth-order free energy with unit cell volume at 300K. Observe the overlap between $F_{2nd}$ and $F_{4th}$ for QHA, resulting from the third-order and fourth-order term cancellation. The inset plot illustrates pressure calculated using the volumetric derivative of free energy.
    (b) Thermal conductivity as a function of pressure at 300K, computed using QHA and renormalized IFCs. Experimental data extracted from Ref\cite{HAKANSSON_Exp_Pressure, Hsieh2021_Pressure}.
    (c) Phonon dispersion of NaCl at 0 GPa and 3 GPa pressure, computed at 300K using renormalized Interatomic Force Constants (IFCs). Observe the significant phonon stiffening, particularly in the optical branches, under increased pressure.  
    (d) Three-phonon scattering rate computed at 0 GPa and 3 GPa pressure. The reduction in anharmonic scattering rate with increased pressure results in higher thermal conductivity at elevated pressures.
    (e) Spectral decomposition of thermal conductivity at 0 GPa and 3 GPa pressure. Notice that the spectral $\kappa$ rises due to the lower scattering rates of phonon modes as pressure increases. 
    (f) and (g) Three-phonon scattering phase space (AAA, AA0, and AOO) of NaCl at two different pressures. Note that there is no significant change in AAA and AOO phase space as pressure increases. The AAO phase space is enhanced for higher-frequency phonon modes due to the widening of the acoustic-optic gap in the phonon dispersion, enabling more higher-frequency modes to participate in scattering events while energy conservation is maintained.}
    \label{fig:Fig_6}
\end{figure*}
\subsection{\label{sec:FreeEng_PVsk} Free energy and pressure dependent thermal conductivity}
Figure \ref{fig:Fig_6}(a) depicts the per unit cell free energy variation as a function of unit cell volume at 300K.
The figure illustrates that the free energy calculated up to the second and fourth order in QHA is nearly identical across all volumes.
In our observations using QHA, we find that the contribution from the third-order and fourth-order anharmonic terms to the free energy mutually nullifies.
Conversely, in the renormalized case, such cancellation does not occur. 
The points of lowest free energy on the curve correspond to the equilibrium lattice constant of the crystal at zero pressure and the specific temperature.
By minimizing the fourth-order free energy using the renormalized IFCs, the lattice constant of NaCl at 300K is determined to be 5.637 $\AA$. The lattice constant obtained from the renormalized harmonic free energy is at 5.624 $\AA$.
In contrast, the QHA predicts the lattice constant to be 5.632 $\AA$.
The experimentally measured lattice constant at NaCl at 300K is 5.645 $\AA$\cite{thermalExpn}. 
While both the QHA and renormalization techniques underpredict the lattice constant, the renormalized fourth-order approach shows better agreement with the measured data.
In Figure \ref{fig:Fig_6}(a), the inset plot illustrates the variation of unit cell volume with pressure for NaCl at 300K.
The pressure ($P$) is determined using the thermodynamic relation involving the free energy: $P = - \left(\frac{\partial F}{\partial V} \right)_T$. 
The pressure-dependent thermal conductivity of NaCl at 300K is presented in Figure \ref{fig:Fig_6}(b). 
As the pressure rises from 0 to 3.4 GPa, the thermal conductivity obtained from QHA experiences an increase, going from 7.53 W/mK to 11.12 W/mK.
As explained in the earlier section, the thermal conductivity obtained from renormalized IFCs tends to overpredict, primarily because we considered only up to three-phonon scattering processes.
The experimental data\cite{HAKANSSON_Exp_Pressure, Hsieh2021_Pressure} also exhibit a similar increasing trend for thermal conductivity, as seen in the QHA values.
The phonon modes stiffen with the increase in pressure, as depicted in Figure \ref{fig:Fig_6}(c). 
It is important to note that the rate of phonon stiffening is more significant in the optic branches compared to the acoustic branches.
In order to comprehend the rise in thermal conductivity with pressure, we calculated three-phonon scattering rates at 0 GPa and 3GPa pressure, as depicted in Figure \ref{fig:Fig_6}(d).
As the pressure increases, the total three-phonon scattering rate decreases, leading to higher thermal conductivity at elevated pressure levels.
The spectral thermal conductivity, as shown in Figure \ref{fig:Fig_6}(e), also shifts to higher values as pressure increases due to reduced scattering rates of phonon modes.
To investigate the anharmonic three-phonon scattering processes further, we calculated the scattering phase space involving different polarizations of phonon modes. 
The scattering phase space considered here is defined as\cite{Ravichandran2019}: 
\begin{eqnarray}\label{eq:PhaseSpace}
    \mathcal{P}(\omega_{qs}) = \frac{2}{3 N_{p}^2 N_0} \sum\limits_{q^{'}, s^{'}} \sum\limits_{q^{''}, s^{''}} \Big[ \delta (\omega_{qs} + \omega_{q^{'} s^{'}} - \omega_{q^{''} s^{''}} ) + \nonumber \\
     \frac{1}{2} \delta (\omega_{qs} - \omega_{q^{'} s^{'}} - \omega_{q^{''} s^{''}} ) \Big],\nonumber \\
\end{eqnarray}
where $N_p$ is the number of polarizations. 
The above expression implicitly incorporates quasi-momentum conservation: $q \pm q^{'} + q^{''} = G$, which leads to considering only the summation over $q^{'}$. 
Multiple scattering processes are feasible, involving various acoustic (A) and optic (O) phonon branch combinations. 
Figures 3(f) and 3(g) illustrate the three-phonon scattering phase space for AAA, AAO, and AOO scattering processes.
Due to energy conservation constraints, scattering processes involving three optical branches (OOO) are not permitted. 
As depicted in Figure \ref{fig:Fig_6}(f), the AAA scattering phase space remains relatively constant as pressure increases. 
AAA scattering phase space at 0 GPa also reveals more feasible scattering events conserving momentum and energy than at 3 GPa.
As depicted in Figure \ref{fig:Fig_6}(g), the AOO scattering phase space demonstrates minimal to no variation with the increase in pressure.
Conversely, the AAO scattering phase space displays an increment with increasing pressure.
This is attributed to the stiffening of phonon dispersion, which results in an increased gap between acoustic and optic phonons. 
According to energy conservation principles, only acoustic phonons with frequencies higher than this are allowed to take part in AAO scattering processes\cite{LinsdeyPhaseSpace, NavneethaPRX}.
So, a greater number of higher frequency phonon modes participate in the scattering process, as evident from Figure \ref{fig:Fig_6}(g).

\section{\label{sec:Conclusions} CONCLUSIONS}
\begin{table*} [!hbt]
	\begin{ruledtabular}
		{\renewcommand{\arraystretch}{1.7}
			\begin{tabular}{|c|l|l|l|l|}
				Material & Temperature (K) & $\kappa$ (QHA) ($W m^{-1}K^{-1}$) & \thead{$\kappa$ ($W m^{-1}K^{-1}$) \\(Renormalized)} & $\kappa$ (Exp.) ($W m^{-1}K^{-1}$)\\ \hline
				& 300K & 648.31 & 666.97 & $\sim$ 648$\pm$52\cite{Exp_kappa_2018}, 850$\pm$90\cite{Exp_kappa_2020}   \\ \cline{2-5}
				cBN     & 400K & 512.95 & 527.42 & $\sim$460$\pm$40\cite{Exp_kappa_2018}, $\sim$603$\pm$70\cite{Exp_kappa_2020}  \\ \cline{2-5}
				& 500K & 423.64 & 437.94 & $\sim$360$\pm$42\cite{Exp_kappa_2018}, $\sim$483$\pm$62\cite{Exp_kappa_2020} \\ 
				\hline
				& 300K & 364.25 & 378.68 & $\sim$437\cite{kappa_singleCrystallineSiC} \\ \cline{2-5}
				3C-SiC  & 200K & 677.71 & 692.97 & $\sim$768\cite{kappa_singleCrystallineSiC} \\ \cline{2-5}
				& 250K & 484.02 & 499.28 & $\sim$552\cite{kappa_singleCrystallineSiC} \\ 
				\hline \hline
				& 300K & 7.53 & 9.95 & 6.02\cite{HAKANSSON_Exp_Pressure}, $\sim$5.8\cite{McCarthy_Exp}, $\sim$8.1\cite{Klein_Exp} \\ \cline{2-5}
				NaCl    & 200K & 10.90 & 13.75 & $\sim$10.27\cite{HAKANSSON_Exp_Pressure}, $\sim$11.35\cite{Klein_Exp} \\ \cline{2-5}
				& 100K & 24.62 & 28.78 & $\sim$22.47\cite{HAKANSSON_Exp_Pressure}, $\sim$22.79\cite{Klein_Exp} \\ 
				\hline
				& 300K & 1.03 & 1.17, \textbf{0.41(4-ph)} & 0.36\cite{AgI_2023_kappa_1}, $\sim$0.41\cite{AgI_Exp_1982} \\ \cline{2-5}
				AgI  & 200K & 1.54 & 1.74, \textbf{0.77(4-ph)} & $\sim$0.61\cite{AgI_2023_kappa_1}, $\sim$0.72\cite{AgI_Exp_1982} \\ \cline{2-5}
				& 400K & 0.78 & 0.88, \textbf{0.25(4-ph)} & $\sim$0.21\cite{AgI_Exp_1982} \\ 
			\end{tabular}
		}
		\caption{\label{Tab:kappa} Comparison of thermal conductivity ($\kappa$) at various temperatures for the materials examined in this study, calculated via the QHA and renormalized IFCs, alongside experimental ($\kappa$) values for reference. The symbol $\sim$ denotes data extracted from corresponding references' figures, which may contain inherent errors.}
	\end{ruledtabular}
\end{table*}
In summary, we have developed a comprehensive numerical framework for understanding phonon and phonon-driven physical properties of crystalline semiconductors and insulators. 
The phonon quasiparticle approach forms the foundation for our self-consistent phonon renormalization technique, which effectively characterizes both the vibrational attributes of strongly anharmonic solids and those of harmonic crystals. Our renormalization technique also extends to the anharmonic third- and fourth-order IFCs.
Both the TDEP and TSS utilized in this study include explicit variations in temperature when computing IFCs.
The investigation of the NaCl crystal in our study demonstrates a notable alignment between the phonon dispersion obtained from the renormalized IFCs and the experimental data.
On the other hand, the thermal conductivity values of NaCl exhibit a close correspondence with the experimental measurements when considering the IFCs obtained through the quasiharmonic approximation (QHA).
Our investigation reveals that both the QHA and the renormalization approach consistently overestimate the experimentally measured thermal conductivity of AgI when considering three-phonon scattering processes. This implies that higher-order phonon scattering processes are crucial considerations in materials with ultralow thermal conductivity, like AgI. Consequently, we undertook the computationally demanding task of calculating fourth-order scattering processes to address these discrepancies.
We observe that materials exhibiting high $\kappa$ experience minimal impact on thermal conductivity and phonon-driven physical properties through the renormalization of the IFCs. 
Both cBN and 3C-SiC, investigated in this study as materials with high thermal conductivity, demonstrate a 2-3\% difference in $\kappa$ values between QHA and renormalization. These results closely match experimental observations, even when considering only up to 3-phonon scattering processes.
In our current investigation, the Helmholtz free energy takes into account anharmonicity up to the fourth order and demonstrates a close agreement with the experimental lattice constant of the NaCl crystal.
Additionally, we have examined how the thermal conductivity of NaCl changes under varying pressure conditions.
Anharmonic three-phonon scattering rates and scattering phase space are also analyzed to elucidate the behavior of the thermal transport coefficient with temperature and pressure. 

While the phonon dispersion and its associated properties exhibit favorable alignment with experimental data through the renormalization approach, anharmonic characteristics like lattice thermal conductivity consistently exhibit an overprediction compared to experimental measurements when IFCs are subjected to renormalization, both for NaCl and AgI. 
The primary reason behind these discrepancies is that we have considered only up to the three-phonon scattering processes. 
The significance of higher-order anharmonicity cannot be overlooked in the case of strongly anharmonic materials such as NaCl and AgI.
While our current exploration focuses solely on the four-phonon scattering processes for AgI, we have omitted the consideration of higher-order scattering processes for other materials in this study due to computational limitations.
The findings of our present investigation are summarized in Table \ref{Tab:kappa}.
The effects of electron-phonon scattering processes\cite{Protik2022-eph} are not accounted for in our current study. 
For certain semiconductors with higher carrier concentrations at standard room temperature, ignoring the influence of electron-phonon scattering processes is not viable.

In conclusion, we offer an extensive numerical framework that enables the exploration of mechanical, thermodynamic, and thermal properties driven by crystalline vibrations in insulators and semiconductors. This framework is built upon a sturdy foundation of interatomic force constant (IFC) calculation and phonon renormalization techniques.
The present development enables the efficient and higher-accuracy numerical and theoretical study for understanding phonon and phonon-driven physical properties for better and more efficient application of materials.

\section{\label{sec:Acknowledgements} ACKNOWLEDGMENTS}
We thank TUE-CMS, IISc, funded by DST, India, as well as Param-Pravega at SERC, IISc, for providing the computational facilities. The authors thank Navaneetha Krishnan Ravichandran for valuable discussions. S.M. acknowledges the generous fellowship provided by MHRD, India. PKM  acknowledges funding through SERB, IRHPA (No. IPA/2020/000034).
\nocite{*}

\bibliography{citations}

\begin{thebibliography}{67}%
\makeatletter
\providecommand \@ifxundefined [1]{%
 \@ifx{#1\undefined}
}%
\providecommand \@ifnum [1]{%
 \ifnum #1\expandafter \@firstoftwo
 \else \expandafter \@secondoftwo
 \fi
}%
\providecommand \@ifx [1]{%
 \ifx #1\expandafter \@firstoftwo
 \else \expandafter \@secondoftwo
 \fi
}%
\providecommand \natexlab [1]{#1}%
\providecommand \enquote  [1]{``#1''}%
\providecommand \bibnamefont  [1]{#1}%
\providecommand \bibfnamefont [1]{#1}%
\providecommand \citenamefont [1]{#1}%
\providecommand \href@noop [0]{\@secondoftwo}%
\providecommand \href [0]{\begingroup \@sanitize@url \@href}%
\providecommand \@href[1]{\@@startlink{#1}\@@href}%
\providecommand \@@href[1]{\endgroup#1\@@endlink}%
\providecommand \@sanitize@url [0]{\catcode `\\12\catcode `\$12\catcode
  `\&12\catcode `\#12\catcode `\^12\catcode `\_12\catcode `\%12\relax}%
\providecommand \@@startlink[1]{}%
\providecommand \@@endlink[0]{}%
\providecommand \url  [0]{\begingroup\@sanitize@url \@url }%
\providecommand \@url [1]{\endgroup\@href {#1}{\urlprefix }}%
\providecommand \urlprefix  [0]{URL }%
\providecommand \Eprint [0]{\href }%
\providecommand \doibase [0]{https://doi.org/}%
\providecommand \selectlanguage [0]{\@gobble}%
\providecommand \bibinfo  [0]{\@secondoftwo}%
\providecommand \bibfield  [0]{\@secondoftwo}%
\providecommand \translation [1]{[#1]}%
\providecommand \BibitemOpen [0]{}%
\providecommand \bibitemStop [0]{}%
\providecommand \bibitemNoStop [0]{.\EOS\space}%
\providecommand \EOS [0]{\spacefactor3000\relax}%
\providecommand \BibitemShut  [1]{\csname bibitem#1\endcsname}%
\let\auto@bib@innerbib\@empty
\bibitem [{\citenamefont {Alfè}(2009)}]{phon_FD}%
  \BibitemOpen
  \bibfield  {author} {\bibinfo {author} {\bibfnamefont {D.}~\bibnamefont
  {Alfè}},\ }\bibfield  {title} {\bibinfo {title} {Phon: A program to
  calculate phonons using the small displacement method},\ }\href
  {https://doi.org/https://doi.org/10.1016/j.cpc.2009.03.010} {\bibfield
  {journal} {\bibinfo  {journal} {Computer Physics Communications}\ }\textbf
  {\bibinfo {volume} {180}},\ \bibinfo {pages} {2622} (\bibinfo {year}
  {2009})}\BibitemShut {NoStop}%
\bibitem [{\citenamefont {Togo}\ and\ \citenamefont
  {Tanaka}(2015)}]{phonopyTOGO2015}%
  \BibitemOpen
  \bibfield  {author} {\bibinfo {author} {\bibfnamefont {A.}~\bibnamefont
  {Togo}}\ and\ \bibinfo {author} {\bibfnamefont {I.}~\bibnamefont {Tanaka}},\
  }\bibfield  {title} {\bibinfo {title} {First principles phonon calculations
  in materials science},\ }\href
  {https://doi.org/https://doi.org/10.1016/j.scriptamat.2015.07.021} {\bibfield
   {journal} {\bibinfo  {journal} {Scripta Materialia}\ }\textbf {\bibinfo
  {volume} {108}},\ \bibinfo {pages} {1} (\bibinfo {year} {2015})}\BibitemShut
  {NoStop}%
\bibitem [{\citenamefont {Togo}\ \emph {et~al.}(2015)\citenamefont {Togo},
  \citenamefont {Chaput},\ and\ \citenamefont {Tanaka}}]{phono3py_2015}%
  \BibitemOpen
  \bibfield  {author} {\bibinfo {author} {\bibfnamefont {A.}~\bibnamefont
  {Togo}}, \bibinfo {author} {\bibfnamefont {L.}~\bibnamefont {Chaput}},\ and\
  \bibinfo {author} {\bibfnamefont {I.}~\bibnamefont {Tanaka}},\ }\bibfield
  {title} {\bibinfo {title} {Distributions of phonon lifetimes in brillouin
  zones},\ }\href {https://doi.org/10.1103/PhysRevB.91.094306} {\bibfield
  {journal} {\bibinfo  {journal} {Phys. Rev. B}\ }\textbf {\bibinfo {volume}
  {91}},\ \bibinfo {pages} {094306} (\bibinfo {year} {2015})}\BibitemShut
  {NoStop}%
\bibitem [{\citenamefont {Li}\ \emph {et~al.}(2014)\citenamefont {Li},
  \citenamefont {Carrete}, \citenamefont {Katcho},\ and\ \citenamefont
  {Mingo}}]{ShengBTE_2014}%
  \BibitemOpen
  \bibfield  {author} {\bibinfo {author} {\bibfnamefont {W.}~\bibnamefont
  {Li}}, \bibinfo {author} {\bibfnamefont {J.}~\bibnamefont {Carrete}},
  \bibinfo {author} {\bibfnamefont {N.~A.}\ \bibnamefont {Katcho}},\ and\
  \bibinfo {author} {\bibfnamefont {N.}~\bibnamefont {Mingo}},\ }\bibfield
  {title} {\bibinfo {title} {{ShengBTE:} a solver of the {B}oltzmann transport
  equation for phonons},\ }\href {https://doi.org/10.1016/j.cpc.2014.02.015}
  {\bibfield  {journal} {\bibinfo  {journal} {Comp. Phys. Commun.}\ }\textbf
  {\bibinfo {volume} {185}},\ \bibinfo {pages} {1747–1758} (\bibinfo {year}
  {2014})}\BibitemShut {NoStop}%
\bibitem [{\citenamefont {Mounet}\ and\ \citenamefont {Marzari}(2005)}]{QHA}%
  \BibitemOpen
  \bibfield  {author} {\bibinfo {author} {\bibfnamefont {N.}~\bibnamefont
  {Mounet}}\ and\ \bibinfo {author} {\bibfnamefont {N.}~\bibnamefont
  {Marzari}},\ }\bibfield  {title} {\bibinfo {title} {First-principles
  determination of the structural, vibrational and thermodynamic properties of
  diamond, graphite, and derivatives},\ }\href
  {https://doi.org/10.1103/PhysRevB.71.205214} {\bibfield  {journal} {\bibinfo
  {journal} {Phys. Rev. B}\ }\textbf {\bibinfo {volume} {71}},\ \bibinfo
  {pages} {205214} (\bibinfo {year} {2005})}\BibitemShut {NoStop}%
\bibitem [{\citenamefont {Wallace}(1972)}]{Wallace}%
  \BibitemOpen
  \bibfield  {author} {\bibinfo {author} {\bibfnamefont {D.~C.}\ \bibnamefont
  {Wallace}},\ }\href@noop {} {\emph {\bibinfo {title} {Thermodynamics of
  Crystals}}}\ (\bibinfo  {publisher} {John Wiley \& Sons, Inc., New York},\
  \bibinfo {year} {1972})\BibitemShut {NoStop}%
\bibitem [{\citenamefont {Baroni}\ \emph {et~al.}(2001)\citenamefont {Baroni},
  \citenamefont {de~Gironcoli}, \citenamefont {Dal~Corso},\ and\ \citenamefont
  {Giannozzi}}]{DFPT}%
  \BibitemOpen
  \bibfield  {author} {\bibinfo {author} {\bibfnamefont {S.}~\bibnamefont
  {Baroni}}, \bibinfo {author} {\bibfnamefont {S.}~\bibnamefont
  {de~Gironcoli}}, \bibinfo {author} {\bibfnamefont {A.}~\bibnamefont
  {Dal~Corso}},\ and\ \bibinfo {author} {\bibfnamefont {P.}~\bibnamefont
  {Giannozzi}},\ }\bibfield  {title} {\bibinfo {title} {Phonons and related
  crystal properties from density-functional perturbation theory},\ }\href
  {https://doi.org/10.1103/RevModPhys.73.515} {\bibfield  {journal} {\bibinfo
  {journal} {Rev. Mod. Phys.}\ }\textbf {\bibinfo {volume} {73}},\ \bibinfo
  {pages} {515} (\bibinfo {year} {2001})}\BibitemShut {NoStop}%
\bibitem [{\citenamefont {Romero}\ \emph {et~al.}(2020)\citenamefont {Romero},
  \citenamefont {Allan}, \citenamefont {Amadon}, \citenamefont {Antonius},
  \citenamefont {Applencourt}, \citenamefont {Baguet}, \citenamefont {Bieder},
  \citenamefont {Bottin}, \citenamefont {Bouchet}, \citenamefont {Bousquet},
  \citenamefont {Bruneval}, \citenamefont {Brunin}, \citenamefont {Caliste},
  \citenamefont {Côté}, \citenamefont {Denier}, \citenamefont {Dreyer},
  \citenamefont {Ghosez}, \citenamefont {Giantomassi}, \citenamefont {Gillet},
  \citenamefont {Gingras}, \citenamefont {Hamann}, \citenamefont {Hautier},
  \citenamefont {Jollet}, \citenamefont {Jomard}, \citenamefont {Martin},
  \citenamefont {Miranda}, \citenamefont {Naccarato}, \citenamefont {Petretto},
  \citenamefont {Pike}, \citenamefont {Planes}, \citenamefont {Prokhorenko},
  \citenamefont {Rangel}, \citenamefont {Ricci}, \citenamefont {Rignanese},
  \citenamefont {Royo}, \citenamefont {Stengel}, \citenamefont {Torrent},
  \citenamefont {van Setten}, \citenamefont {Van~Troeye}, \citenamefont
  {Verstraete}, \citenamefont {Wiktor}, \citenamefont {Zwanziger},\ and\
  \citenamefont {Gonze}}]{abinit_DFPT}%
  \BibitemOpen
  \bibfield  {author} {\bibinfo {author} {\bibfnamefont {A.~H.}\ \bibnamefont
  {Romero}}, \bibinfo {author} {\bibfnamefont {D.~C.}\ \bibnamefont {Allan}},
  \bibinfo {author} {\bibfnamefont {B.}~\bibnamefont {Amadon}}, \bibinfo
  {author} {\bibfnamefont {G.}~\bibnamefont {Antonius}}, \bibinfo {author}
  {\bibfnamefont {T.}~\bibnamefont {Applencourt}}, \bibinfo {author}
  {\bibfnamefont {L.}~\bibnamefont {Baguet}}, \bibinfo {author} {\bibfnamefont
  {J.}~\bibnamefont {Bieder}}, \bibinfo {author} {\bibfnamefont
  {F.}~\bibnamefont {Bottin}}, \bibinfo {author} {\bibfnamefont
  {J.}~\bibnamefont {Bouchet}}, \bibinfo {author} {\bibfnamefont
  {E.}~\bibnamefont {Bousquet}}, \bibinfo {author} {\bibfnamefont
  {F.}~\bibnamefont {Bruneval}}, \bibinfo {author} {\bibfnamefont
  {G.}~\bibnamefont {Brunin}}, \bibinfo {author} {\bibfnamefont
  {D.}~\bibnamefont {Caliste}}, \bibinfo {author} {\bibfnamefont
  {M.}~\bibnamefont {Côté}}, \bibinfo {author} {\bibfnamefont
  {J.}~\bibnamefont {Denier}}, \bibinfo {author} {\bibfnamefont
  {C.}~\bibnamefont {Dreyer}}, \bibinfo {author} {\bibfnamefont
  {P.}~\bibnamefont {Ghosez}}, \bibinfo {author} {\bibfnamefont
  {M.}~\bibnamefont {Giantomassi}}, \bibinfo {author} {\bibfnamefont
  {Y.}~\bibnamefont {Gillet}}, \bibinfo {author} {\bibfnamefont
  {O.}~\bibnamefont {Gingras}}, \bibinfo {author} {\bibfnamefont {D.~R.}\
  \bibnamefont {Hamann}}, \bibinfo {author} {\bibfnamefont {G.}~\bibnamefont
  {Hautier}}, \bibinfo {author} {\bibfnamefont {F.}~\bibnamefont {Jollet}},
  \bibinfo {author} {\bibfnamefont {G.}~\bibnamefont {Jomard}}, \bibinfo
  {author} {\bibfnamefont {A.}~\bibnamefont {Martin}}, \bibinfo {author}
  {\bibfnamefont {H.~P.~C.}\ \bibnamefont {Miranda}}, \bibinfo {author}
  {\bibfnamefont {F.}~\bibnamefont {Naccarato}}, \bibinfo {author}
  {\bibfnamefont {G.}~\bibnamefont {Petretto}}, \bibinfo {author}
  {\bibfnamefont {N.~A.}\ \bibnamefont {Pike}}, \bibinfo {author}
  {\bibfnamefont {V.}~\bibnamefont {Planes}}, \bibinfo {author} {\bibfnamefont
  {S.}~\bibnamefont {Prokhorenko}}, \bibinfo {author} {\bibfnamefont
  {T.}~\bibnamefont {Rangel}}, \bibinfo {author} {\bibfnamefont
  {F.}~\bibnamefont {Ricci}}, \bibinfo {author} {\bibfnamefont {G.-M.}\
  \bibnamefont {Rignanese}}, \bibinfo {author} {\bibfnamefont {M.}~\bibnamefont
  {Royo}}, \bibinfo {author} {\bibfnamefont {M.}~\bibnamefont {Stengel}},
  \bibinfo {author} {\bibfnamefont {M.}~\bibnamefont {Torrent}}, \bibinfo
  {author} {\bibfnamefont {M.~J.}\ \bibnamefont {van Setten}}, \bibinfo
  {author} {\bibfnamefont {B.}~\bibnamefont {Van~Troeye}}, \bibinfo {author}
  {\bibfnamefont {M.~J.}\ \bibnamefont {Verstraete}}, \bibinfo {author}
  {\bibfnamefont {J.}~\bibnamefont {Wiktor}}, \bibinfo {author} {\bibfnamefont
  {J.~W.}\ \bibnamefont {Zwanziger}},\ and\ \bibinfo {author} {\bibfnamefont
  {X.}~\bibnamefont {Gonze}},\ }\bibfield  {title} {\bibinfo {title} {{ABINIT:
  Overview and focus on selected capabilities}},\ }\href
  {https://doi.org/10.1063/1.5144261} {\bibfield  {journal} {\bibinfo
  {journal} {The Journal of Chemical Physics}\ }\textbf {\bibinfo {volume}
  {152}},\ \bibinfo {pages} {124102} (\bibinfo {year} {2020})},\ \Eprint
  {https://arxiv.org/abs/https://pubs.aip.org/aip/jcp/article-pdf/doi/10.1063/1.5144261/16711978/124102\_1\_online.pdf}
  {https://pubs.aip.org/aip/jcp/article-pdf/doi/10.1063/1.5144261/16711978/124102\_1\_online.pdf}
  \BibitemShut {NoStop}%
\bibitem [{\citenamefont {Li}\ \emph {et~al.}(2012)\citenamefont {Li},
  \citenamefont {Lindsay}, \citenamefont {Broido}, \citenamefont {Stewart},\
  and\ \citenamefont {Mingo}}]{IFC_symm}%
  \BibitemOpen
  \bibfield  {author} {\bibinfo {author} {\bibfnamefont {W.}~\bibnamefont
  {Li}}, \bibinfo {author} {\bibfnamefont {L.}~\bibnamefont {Lindsay}},
  \bibinfo {author} {\bibfnamefont {D.~A.}\ \bibnamefont {Broido}}, \bibinfo
  {author} {\bibfnamefont {D.~A.}\ \bibnamefont {Stewart}},\ and\ \bibinfo
  {author} {\bibfnamefont {N.}~\bibnamefont {Mingo}},\ }\bibfield  {title}
  {\bibinfo {title} {Thermal conductivity of bulk and nanowire
  mg${}_{2}$si${}_{x}$sn${}_{1\ensuremath{-}x}$ alloys from first principles},\
  }\href {https://doi.org/10.1103/PhysRevB.86.174307} {\bibfield  {journal}
  {\bibinfo  {journal} {Phys. Rev. B}\ }\textbf {\bibinfo {volume} {86}},\
  \bibinfo {pages} {174307} (\bibinfo {year} {2012})}\BibitemShut {NoStop}%
\bibitem [{\citenamefont {Togo}\ \emph {et~al.}(2023)\citenamefont {Togo},
  \citenamefont {Chaput}, \citenamefont {Tadano},\ and\ \citenamefont
  {Tanaka}}]{phonopy-phono3py-JPCM}%
  \BibitemOpen
  \bibfield  {author} {\bibinfo {author} {\bibfnamefont {A.}~\bibnamefont
  {Togo}}, \bibinfo {author} {\bibfnamefont {L.}~\bibnamefont {Chaput}},
  \bibinfo {author} {\bibfnamefont {T.}~\bibnamefont {Tadano}},\ and\ \bibinfo
  {author} {\bibfnamefont {I.}~\bibnamefont {Tanaka}},\ }\bibfield  {title}
  {\bibinfo {title} {Implementation strategies in phonopy and phono3py},\
  }\href {https://doi.org/10.1088/1361-648X/acd831} {\bibfield  {journal}
  {\bibinfo  {journal} {J. Phys. Condens. Matter}\ }\textbf {\bibinfo {volume}
  {35}},\ \bibinfo {pages} {353001} (\bibinfo {year} {2023})}\BibitemShut
  {NoStop}%
\bibitem [{\citenamefont {Togo}(2023)}]{phonopy-phono3py-JPSJ}%
  \BibitemOpen
  \bibfield  {author} {\bibinfo {author} {\bibfnamefont {A.}~\bibnamefont
  {Togo}},\ }\bibfield  {title} {\bibinfo {title} {First-principles phonon
  calculations with phonopy and phono3py},\ }\href
  {https://doi.org/10.7566/JPSJ.92.012001} {\bibfield  {journal} {\bibinfo
  {journal} {J. Phys. Soc. Jpn.}\ }\textbf {\bibinfo {volume} {92}},\ \bibinfo
  {pages} {012001} (\bibinfo {year} {2023})}\BibitemShut {NoStop}%
\bibitem [{\citenamefont {Paulatto}\ \emph {et~al.}(2013)\citenamefont
  {Paulatto}, \citenamefont {Mauri},\ and\ \citenamefont {Lazzeri}}]{D3Q}%
  \BibitemOpen
  \bibfield  {author} {\bibinfo {author} {\bibfnamefont {L.}~\bibnamefont
  {Paulatto}}, \bibinfo {author} {\bibfnamefont {F.}~\bibnamefont {Mauri}},\
  and\ \bibinfo {author} {\bibfnamefont {M.}~\bibnamefont {Lazzeri}},\
  }\bibfield  {title} {\bibinfo {title} {Anharmonic properties from a
  generalized third-order ab initio approach: Theory and applications to
  graphite and graphene},\ }\href {https://doi.org/10.1103/PhysRevB.87.214303}
  {\bibfield  {journal} {\bibinfo  {journal} {Phys. Rev. B}\ }\textbf {\bibinfo
  {volume} {87}},\ \bibinfo {pages} {214303} (\bibinfo {year}
  {2013})}\BibitemShut {NoStop}%
\bibitem [{\citenamefont {Han}\ \emph {et~al.}(2022)\citenamefont {Han},
  \citenamefont {Yang}, \citenamefont {Li}, \citenamefont {Feng},\ and\
  \citenamefont {Ruan}}]{FourPhonon}%
  \BibitemOpen
  \bibfield  {author} {\bibinfo {author} {\bibfnamefont {Z.}~\bibnamefont
  {Han}}, \bibinfo {author} {\bibfnamefont {X.}~\bibnamefont {Yang}}, \bibinfo
  {author} {\bibfnamefont {W.}~\bibnamefont {Li}}, \bibinfo {author}
  {\bibfnamefont {T.}~\bibnamefont {Feng}},\ and\ \bibinfo {author}
  {\bibfnamefont {X.}~\bibnamefont {Ruan}},\ }\bibfield  {title} {\bibinfo
  {title} {Fourphonon: An extension module to shengbte for computing
  four-phonon scattering rates and thermal conductivity},\ }\href
  {https://doi.org/https://doi.org/10.1016/j.cpc.2021.108179} {\bibfield
  {journal} {\bibinfo  {journal} {Computer Physics Communications}\ }\textbf
  {\bibinfo {volume} {270}},\ \bibinfo {pages} {108179} (\bibinfo {year}
  {2022})}\BibitemShut {NoStop}%
\bibitem [{\citenamefont {Feng}\ and\ \citenamefont
  {Ruan}(2016)}]{Ruan4ph_2016}%
  \BibitemOpen
  \bibfield  {author} {\bibinfo {author} {\bibfnamefont {T.}~\bibnamefont
  {Feng}}\ and\ \bibinfo {author} {\bibfnamefont {X.}~\bibnamefont {Ruan}},\
  }\bibfield  {title} {\bibinfo {title} {Quantum mechanical prediction of
  four-phonon scattering rates and reduced thermal conductivity of solids},\
  }\href {https://doi.org/10.1103/PhysRevB.93.045202} {\bibfield  {journal}
  {\bibinfo  {journal} {Phys. Rev. B}\ }\textbf {\bibinfo {volume} {93}},\
  \bibinfo {pages} {045202} (\bibinfo {year} {2016})}\BibitemShut {NoStop}%
\bibitem [{\citenamefont {Feng}\ \emph {et~al.}(2017)\citenamefont {Feng},
  \citenamefont {Lindsay},\ and\ \citenamefont {Ruan}}]{Ruan4ph_2017}%
  \BibitemOpen
  \bibfield  {author} {\bibinfo {author} {\bibfnamefont {T.}~\bibnamefont
  {Feng}}, \bibinfo {author} {\bibfnamefont {L.}~\bibnamefont {Lindsay}},\ and\
  \bibinfo {author} {\bibfnamefont {X.}~\bibnamefont {Ruan}},\ }\bibfield
  {title} {\bibinfo {title} {Four-phonon scattering significantly reduces
  intrinsic thermal conductivity of solids},\ }\href
  {https://doi.org/10.1103/PhysRevB.96.161201} {\bibfield  {journal} {\bibinfo
  {journal} {Phys. Rev. B}\ }\textbf {\bibinfo {volume} {96}},\ \bibinfo
  {pages} {161201} (\bibinfo {year} {2017})}\BibitemShut {NoStop}%
\bibitem [{\citenamefont {Han}\ and\ \citenamefont {Ruan}(2023)}]{Ruan_4ph}%
  \BibitemOpen
  \bibfield  {author} {\bibinfo {author} {\bibfnamefont {Z.}~\bibnamefont
  {Han}}\ and\ \bibinfo {author} {\bibfnamefont {X.}~\bibnamefont {Ruan}},\
  }\bibfield  {title} {\bibinfo {title} {Thermal conductivity of monolayer
  graphene: Convergent and lower than diamond},\ }\href
  {https://doi.org/10.1103/PhysRevB.108.L121412} {\bibfield  {journal}
  {\bibinfo  {journal} {Phys. Rev. B}\ }\textbf {\bibinfo {volume} {108}},\
  \bibinfo {pages} {L121412} (\bibinfo {year} {2023})}\BibitemShut {NoStop}%
\bibitem [{\citenamefont {Xia}\ \emph {et~al.}(2020)\citenamefont {Xia},
  \citenamefont {Hegde}, \citenamefont {Pal}, \citenamefont {Hua},
  \citenamefont {Gaines}, \citenamefont {Patel}, \citenamefont {He},
  \citenamefont {Aykol},\ and\ \citenamefont {Wolverton}}]{4-ph_KPal}%
  \BibitemOpen
  \bibfield  {author} {\bibinfo {author} {\bibfnamefont {Y.}~\bibnamefont
  {Xia}}, \bibinfo {author} {\bibfnamefont {V.~I.}\ \bibnamefont {Hegde}},
  \bibinfo {author} {\bibfnamefont {K.}~\bibnamefont {Pal}}, \bibinfo {author}
  {\bibfnamefont {X.}~\bibnamefont {Hua}}, \bibinfo {author} {\bibfnamefont
  {D.}~\bibnamefont {Gaines}}, \bibinfo {author} {\bibfnamefont
  {S.}~\bibnamefont {Patel}}, \bibinfo {author} {\bibfnamefont
  {J.}~\bibnamefont {He}}, \bibinfo {author} {\bibfnamefont {M.}~\bibnamefont
  {Aykol}},\ and\ \bibinfo {author} {\bibfnamefont {C.}~\bibnamefont
  {Wolverton}},\ }\bibfield  {title} {\bibinfo {title} {High-throughput study
  of lattice thermal conductivity in binary rocksalt and zinc blende compounds
  including higher-order anharmonicity},\ }\href
  {https://doi.org/10.1103/PhysRevX.10.041029} {\bibfield  {journal} {\bibinfo
  {journal} {Phys. Rev. X}\ }\textbf {\bibinfo {volume} {10}},\ \bibinfo
  {pages} {041029} (\bibinfo {year} {2020})}\BibitemShut {NoStop}%
\bibitem [{\citenamefont {Knoop}\ \emph
  {et~al.}(2023{\natexlab{a}})\citenamefont {Knoop}, \citenamefont {Purcell},
  \citenamefont {Scheffler},\ and\ \citenamefont {Carbogno}}]{GK_prl}%
  \BibitemOpen
  \bibfield  {author} {\bibinfo {author} {\bibfnamefont {F.}~\bibnamefont
  {Knoop}}, \bibinfo {author} {\bibfnamefont {T.~A.~R.}\ \bibnamefont
  {Purcell}}, \bibinfo {author} {\bibfnamefont {M.}~\bibnamefont {Scheffler}},\
  and\ \bibinfo {author} {\bibfnamefont {C.}~\bibnamefont {Carbogno}},\
  }\bibfield  {title} {\bibinfo {title} {Anharmonicity in thermal insulators:
  An analysis from first principles},\ }\href
  {https://doi.org/10.1103/PhysRevLett.130.236301} {\bibfield  {journal}
  {\bibinfo  {journal} {Phys. Rev. Lett.}\ }\textbf {\bibinfo {volume} {130}},\
  \bibinfo {pages} {236301} (\bibinfo {year} {2023}{\natexlab{a}})}\BibitemShut
  {NoStop}%
\bibitem [{\citenamefont {Knoop}\ \emph
  {et~al.}(2023{\natexlab{b}})\citenamefont {Knoop}, \citenamefont
  {Scheffler},\ and\ \citenamefont {Carbogno}}]{GK_prb}%
  \BibitemOpen
  \bibfield  {author} {\bibinfo {author} {\bibfnamefont {F.}~\bibnamefont
  {Knoop}}, \bibinfo {author} {\bibfnamefont {M.}~\bibnamefont {Scheffler}},\
  and\ \bibinfo {author} {\bibfnamefont {C.}~\bibnamefont {Carbogno}},\
  }\bibfield  {title} {\bibinfo {title} {Ab initio green-kubo simulations of
  heat transport in solids: Method and implementation},\ }\href
  {https://doi.org/10.1103/PhysRevB.107.224304} {\bibfield  {journal} {\bibinfo
   {journal} {Phys. Rev. B}\ }\textbf {\bibinfo {volume} {107}},\ \bibinfo
  {pages} {224304} (\bibinfo {year} {2023}{\natexlab{b}})}\BibitemShut
  {NoStop}%
\bibitem [{\citenamefont {Mandal}\ \emph {et~al.}(2022)\citenamefont {Mandal},
  \citenamefont {Maity}, \citenamefont {Das}, \citenamefont {Jain},\ and\
  \citenamefont {Maiti}}]{PCCP}%
  \BibitemOpen
  \bibfield  {author} {\bibinfo {author} {\bibfnamefont {S.}~\bibnamefont
  {Mandal}}, \bibinfo {author} {\bibfnamefont {I.}~\bibnamefont {Maity}},
  \bibinfo {author} {\bibfnamefont {A.}~\bibnamefont {Das}}, \bibinfo {author}
  {\bibfnamefont {M.}~\bibnamefont {Jain}},\ and\ \bibinfo {author}
  {\bibfnamefont {P.~K.}\ \bibnamefont {Maiti}},\ }\bibfield  {title} {\bibinfo
  {title} {Tunable lattice thermal conductivity of twisted bilayer mos2},\
  }\href {https://doi.org/10.1039/D2CP01304E} {\bibfield  {journal} {\bibinfo
  {journal} {Phys. Chem. Chem. Phys.}\ }\textbf {\bibinfo {volume} {24}},\
  \bibinfo {pages} {13860} (\bibinfo {year} {2022})}\BibitemShut {NoStop}%
\bibitem [{\citenamefont {Hellman}\ \emph {et~al.}(2011)\citenamefont
  {Hellman}, \citenamefont {Abrikosov},\ and\ \citenamefont
  {Simak}}]{TDEP_2011}%
  \BibitemOpen
  \bibfield  {author} {\bibinfo {author} {\bibfnamefont {O.}~\bibnamefont
  {Hellman}}, \bibinfo {author} {\bibfnamefont {I.~A.}\ \bibnamefont
  {Abrikosov}},\ and\ \bibinfo {author} {\bibfnamefont {S.~I.}\ \bibnamefont
  {Simak}},\ }\bibfield  {title} {\bibinfo {title} {Lattice dynamics of
  anharmonic solids from first principles},\ }\href
  {https://doi.org/10.1103/PhysRevB.84.180301} {\bibfield  {journal} {\bibinfo
  {journal} {Phys. Rev. B}\ }\textbf {\bibinfo {volume} {84}},\ \bibinfo
  {pages} {180301} (\bibinfo {year} {2011})}\BibitemShut {NoStop}%
\bibitem [{\citenamefont {Hellman}\ and\ \citenamefont
  {Abrikosov}(2013)}]{TDEP_3rd}%
  \BibitemOpen
  \bibfield  {author} {\bibinfo {author} {\bibfnamefont {O.}~\bibnamefont
  {Hellman}}\ and\ \bibinfo {author} {\bibfnamefont {I.~A.}\ \bibnamefont
  {Abrikosov}},\ }\bibfield  {title} {\bibinfo {title} {Temperature-dependent
  effective third-order interatomic force constants from first principles},\
  }\href {https://doi.org/10.1103/PhysRevB.88.144301} {\bibfield  {journal}
  {\bibinfo  {journal} {Phys. Rev. B}\ }\textbf {\bibinfo {volume} {88}},\
  \bibinfo {pages} {144301} (\bibinfo {year} {2013})}\BibitemShut {NoStop}%
\bibitem [{\citenamefont {Bottin}\ \emph {et~al.}(2020)\citenamefont {Bottin},
  \citenamefont {Bieder},\ and\ \citenamefont {Bouchet}}]{a-TDEP}%
  \BibitemOpen
  \bibfield  {author} {\bibinfo {author} {\bibfnamefont {F.}~\bibnamefont
  {Bottin}}, \bibinfo {author} {\bibfnamefont {J.}~\bibnamefont {Bieder}},\
  and\ \bibinfo {author} {\bibfnamefont {J.}~\bibnamefont {Bouchet}},\
  }\bibfield  {title} {\bibinfo {title} {a-tdep: Temperature dependent
  effective potential for abinit – lattice dynamic properties including
  anharmonicity},\ }\href
  {https://doi.org/https://doi.org/10.1016/j.cpc.2020.107301} {\bibfield
  {journal} {\bibinfo  {journal} {Computer Physics Communications}\ }\textbf
  {\bibinfo {volume} {254}},\ \bibinfo {pages} {107301} (\bibinfo {year}
  {2020})}\BibitemShut {NoStop}%
\bibitem [{\citenamefont {Jain}(2020)}]{AnkitJain_2020}%
  \BibitemOpen
  \bibfield  {author} {\bibinfo {author} {\bibfnamefont {A.}~\bibnamefont
  {Jain}},\ }\bibfield  {title} {\bibinfo {title} {Multichannel thermal
  transport in crystalline ${\mathrm{tl}}_{3}{\mathrm{vse}}_{4}$},\ }\href
  {https://doi.org/10.1103/PhysRevB.102.201201} {\bibfield  {journal} {\bibinfo
   {journal} {Phys. Rev. B}\ }\textbf {\bibinfo {volume} {102}},\ \bibinfo
  {pages} {201201} (\bibinfo {year} {2020})}\BibitemShut {NoStop}%
\bibitem [{\citenamefont {Tadano}\ \emph {et~al.}(2014)\citenamefont {Tadano},
  \citenamefont {Gohda},\ and\ \citenamefont {Tsuneyuki}}]{alamode_2014}%
  \BibitemOpen
  \bibfield  {author} {\bibinfo {author} {\bibfnamefont {T.}~\bibnamefont
  {Tadano}}, \bibinfo {author} {\bibfnamefont {Y.}~\bibnamefont {Gohda}},\ and\
  \bibinfo {author} {\bibfnamefont {S.}~\bibnamefont {Tsuneyuki}},\ }\bibfield
  {title} {\bibinfo {title} {Anharmonic force constants extracted from
  first-principles molecular dynamics: applications to heat transfer
  simulations},\ }\href {https://doi.org/10.1088/0953-8984/26/22/225402}
  {\bibfield  {journal} {\bibinfo  {journal} {Journal of Physics: Condensed
  Matter}\ }\textbf {\bibinfo {volume} {26}},\ \bibinfo {pages} {225402}
  (\bibinfo {year} {2014})}\BibitemShut {NoStop}%
\bibitem [{\citenamefont {Mahan}(2000)}]{GDMahan}%
  \BibitemOpen
  \bibfield  {author} {\bibinfo {author} {\bibfnamefont {G.~D.}\ \bibnamefont
  {Mahan}},\ }\href {https://doi.org/https://doi.org/10.1007/978-1-4757-5714-9}
  {\emph {\bibinfo {title} {Many-Particle Physics}}}\ (\bibinfo  {publisher}
  {Springer New York, NY},\ \bibinfo {year} {2000})\BibitemShut {NoStop}%
\bibitem [{\citenamefont {Anderson}\ \emph {et~al.}(1999)\citenamefont
  {Anderson}, \citenamefont {Bai}, \citenamefont {Bischof}, \citenamefont
  {Blackford}, \citenamefont {Demmel}, \citenamefont {Dongarra}, \citenamefont
  {Du~Croz}, \citenamefont {Greenbaum}, \citenamefont {Hammarling},
  \citenamefont {McKenney},\ and\ \citenamefont {Sorensen}}]{laug}%
  \BibitemOpen
  \bibfield  {author} {\bibinfo {author} {\bibfnamefont {E.}~\bibnamefont
  {Anderson}}, \bibinfo {author} {\bibfnamefont {Z.}~\bibnamefont {Bai}},
  \bibinfo {author} {\bibfnamefont {C.}~\bibnamefont {Bischof}}, \bibinfo
  {author} {\bibfnamefont {S.}~\bibnamefont {Blackford}}, \bibinfo {author}
  {\bibfnamefont {J.}~\bibnamefont {Demmel}}, \bibinfo {author} {\bibfnamefont
  {J.}~\bibnamefont {Dongarra}}, \bibinfo {author} {\bibfnamefont
  {J.}~\bibnamefont {Du~Croz}}, \bibinfo {author} {\bibfnamefont
  {A.}~\bibnamefont {Greenbaum}}, \bibinfo {author} {\bibfnamefont
  {S.}~\bibnamefont {Hammarling}}, \bibinfo {author} {\bibfnamefont
  {A.}~\bibnamefont {McKenney}},\ and\ \bibinfo {author} {\bibfnamefont
  {D.}~\bibnamefont {Sorensen}},\ }\href@noop {} {\emph {\bibinfo {title}
  {{LAPACK} Users' Guide}}},\ \bibinfo {edition} {3rd}\ ed.\ (\bibinfo
  {publisher} {Society for Industrial and Applied Mathematics},\ \bibinfo
  {address} {Philadelphia, PA},\ \bibinfo {year} {1999})\BibitemShut {NoStop}%
\bibitem [{\citenamefont {Togo}\ and\ \citenamefont {Tanaka}(2018)}]{Spglib}%
  \BibitemOpen
  \bibfield  {author} {\bibinfo {author} {\bibfnamefont {A.}~\bibnamefont
  {Togo}}\ and\ \bibinfo {author} {\bibfnamefont {I.}~\bibnamefont {Tanaka}},\
  }\bibfield  {title} {\bibinfo {title} {{$\texttt{Spglib}$}: a software
  library for crystal symmetry search}\ }\href
  {https://doi.org/10.48550/arXiv.1808.01590} {10.48550/arXiv.1808.01590}
  (\bibinfo {year} {2018}),\ \Eprint {https://arxiv.org/abs/1808.01590}
  {arXiv:1808.01590 [cond-mat.mtrl-sci]} \BibitemShut {NoStop}%
\bibitem [{\citenamefont {Wang}\ \emph {et~al.}(2010)\citenamefont {Wang},
  \citenamefont {Wang}, \citenamefont {Wang}, \citenamefont {Mei},
  \citenamefont {Shang}, \citenamefont {Chen},\ and\ \citenamefont
  {Liu}}]{mixed_space_2010}%
  \BibitemOpen
  \bibfield  {author} {\bibinfo {author} {\bibfnamefont {Y.}~\bibnamefont
  {Wang}}, \bibinfo {author} {\bibfnamefont {J.~J.}\ \bibnamefont {Wang}},
  \bibinfo {author} {\bibfnamefont {W.~Y.}\ \bibnamefont {Wang}}, \bibinfo
  {author} {\bibfnamefont {Z.~G.}\ \bibnamefont {Mei}}, \bibinfo {author}
  {\bibfnamefont {S.~L.}\ \bibnamefont {Shang}}, \bibinfo {author}
  {\bibfnamefont {L.~Q.}\ \bibnamefont {Chen}},\ and\ \bibinfo {author}
  {\bibfnamefont {Z.~K.}\ \bibnamefont {Liu}},\ }\bibfield  {title} {\bibinfo
  {title} {A mixed-space approach to first-principles calculations of phonon
  frequencies for polar materials},\ }\href
  {https://doi.org/10.1088/0953-8984/22/20/202201} {\bibfield  {journal}
  {\bibinfo  {journal} {Journal of Physics: Condensed Matter}\ }\textbf
  {\bibinfo {volume} {22}},\ \bibinfo {pages} {202201} (\bibinfo {year}
  {2010})}\BibitemShut {NoStop}%
\bibitem [{\citenamefont {Gonze}\ \emph {et~al.}(1994)\citenamefont {Gonze},
  \citenamefont {Charlier}, \citenamefont {Allan},\ and\ \citenamefont
  {Teter}}]{EwaldNonAnalytic}%
  \BibitemOpen
  \bibfield  {author} {\bibinfo {author} {\bibfnamefont {X.}~\bibnamefont
  {Gonze}}, \bibinfo {author} {\bibfnamefont {J.-C.}\ \bibnamefont {Charlier}},
  \bibinfo {author} {\bibfnamefont {D.}~\bibnamefont {Allan}},\ and\ \bibinfo
  {author} {\bibfnamefont {M.}~\bibnamefont {Teter}},\ }\bibfield  {title}
  {\bibinfo {title} {Interatomic force constants from first principles: The
  case of \ensuremath{\alpha}-quartz},\ }\href
  {https://doi.org/10.1103/PhysRevB.50.13035} {\bibfield  {journal} {\bibinfo
  {journal} {Phys. Rev. B}\ }\textbf {\bibinfo {volume} {50}},\ \bibinfo
  {pages} {13035} (\bibinfo {year} {1994})}\BibitemShut {NoStop}%
\bibitem [{\citenamefont {Gonze}\ and\ \citenamefont
  {Lee}(1997)}]{EwaldNonAnalytic2}%
  \BibitemOpen
  \bibfield  {author} {\bibinfo {author} {\bibfnamefont {X.}~\bibnamefont
  {Gonze}}\ and\ \bibinfo {author} {\bibfnamefont {C.}~\bibnamefont {Lee}},\
  }\bibfield  {title} {\bibinfo {title} {Dynamical matrices, born effective
  charges, dielectric permittivity tensors, and interatomic force constants
  from density-functional perturbation theory},\ }\href
  {https://doi.org/10.1103/PhysRevB.55.10355} {\bibfield  {journal} {\bibinfo
  {journal} {Phys. Rev. B}\ }\textbf {\bibinfo {volume} {55}},\ \bibinfo
  {pages} {10355} (\bibinfo {year} {1997})}\BibitemShut {NoStop}%
\bibitem [{\citenamefont {West}\ and\ \citenamefont
  {Estreicher}(2006)}]{TSS_2006}%
  \BibitemOpen
  \bibfield  {author} {\bibinfo {author} {\bibfnamefont {D.}~\bibnamefont
  {West}}\ and\ \bibinfo {author} {\bibfnamefont {S.~K.}\ \bibnamefont
  {Estreicher}},\ }\bibfield  {title} {\bibinfo {title} {First-principles
  calculations of vibrational lifetimes and decay channels: Hydrogen-related
  modes in si},\ }\href {https://doi.org/10.1103/PhysRevLett.96.115504}
  {\bibfield  {journal} {\bibinfo  {journal} {Phys. Rev. Lett.}\ }\textbf
  {\bibinfo {volume} {96}},\ \bibinfo {pages} {115504} (\bibinfo {year}
  {2006})}\BibitemShut {NoStop}%
\bibitem [{\citenamefont {Shulumba}\ \emph {et~al.}(2017)\citenamefont
  {Shulumba}, \citenamefont {Hellman},\ and\ \citenamefont
  {Minnich}}]{TDEP_TSS}%
  \BibitemOpen
  \bibfield  {author} {\bibinfo {author} {\bibfnamefont {N.}~\bibnamefont
  {Shulumba}}, \bibinfo {author} {\bibfnamefont {O.}~\bibnamefont {Hellman}},\
  and\ \bibinfo {author} {\bibfnamefont {A.~J.}\ \bibnamefont {Minnich}},\
  }\bibfield  {title} {\bibinfo {title} {Intrinsic localized mode and low
  thermal conductivity of pbse},\ }\href
  {https://doi.org/10.1103/PhysRevB.95.014302} {\bibfield  {journal} {\bibinfo
  {journal} {Phys. Rev. B}\ }\textbf {\bibinfo {volume} {95}},\ \bibinfo
  {pages} {014302} (\bibinfo {year} {2017})}\BibitemShut {NoStop}%
\bibitem [{\citenamefont {Ravichandran}\ and\ \citenamefont
  {Broido}(2018)}]{Ravichandran_2018}%
  \BibitemOpen
  \bibfield  {author} {\bibinfo {author} {\bibfnamefont {N.~K.}\ \bibnamefont
  {Ravichandran}}\ and\ \bibinfo {author} {\bibfnamefont {D.}~\bibnamefont
  {Broido}},\ }\bibfield  {title} {\bibinfo {title} {Unified first-principles
  theory of thermal properties of insulators},\ }\href
  {https://doi.org/10.1103/PhysRevB.98.085205} {\bibfield  {journal} {\bibinfo
  {journal} {Phys. Rev. B}\ }\textbf {\bibinfo {volume} {98}},\ \bibinfo
  {pages} {085205} (\bibinfo {year} {2018})}\BibitemShut {NoStop}%
\bibitem [{\citenamefont {Soler}\ \emph {et~al.}(2002)\citenamefont {Soler},
  \citenamefont {Artacho}, \citenamefont {Gale}, \citenamefont {García},
  \citenamefont {Junquera}, \citenamefont {Ordejón},\ and\ \citenamefont
  {Sánchez-Portal}}]{SIESTA_2002}%
  \BibitemOpen
  \bibfield  {author} {\bibinfo {author} {\bibfnamefont {J.~M.}\ \bibnamefont
  {Soler}}, \bibinfo {author} {\bibfnamefont {E.}~\bibnamefont {Artacho}},
  \bibinfo {author} {\bibfnamefont {J.~D.}\ \bibnamefont {Gale}}, \bibinfo
  {author} {\bibfnamefont {A.}~\bibnamefont {García}}, \bibinfo {author}
  {\bibfnamefont {J.}~\bibnamefont {Junquera}}, \bibinfo {author}
  {\bibfnamefont {P.}~\bibnamefont {Ordejón}},\ and\ \bibinfo {author}
  {\bibfnamefont {D.}~\bibnamefont {Sánchez-Portal}},\ }\bibfield  {title}
  {\bibinfo {title} {The siesta method for ab initio order-n materials
  simulation},\ }\href {https://doi.org/10.1088/0953-8984/14/11/302} {\bibfield
   {journal} {\bibinfo  {journal} {Journal of Physics: Condensed Matter}\
  }\textbf {\bibinfo {volume} {14}},\ \bibinfo {pages} {2745} (\bibinfo {year}
  {2002})}\BibitemShut {NoStop}%
\bibitem [{\citenamefont {Pal}\ \emph {et~al.}(2021)\citenamefont {Pal},
  \citenamefont {Xia},\ and\ \citenamefont {Wolverton}}]{Ren_Kpal}%
  \BibitemOpen
  \bibfield  {author} {\bibinfo {author} {\bibfnamefont {K.}~\bibnamefont
  {Pal}}, \bibinfo {author} {\bibfnamefont {Y.}~\bibnamefont {Xia}},\ and\
  \bibinfo {author} {\bibfnamefont {C.}~\bibnamefont {Wolverton}},\ }\bibfield
  {title} {\bibinfo {title} {Microscopic mechanism of unusual lattice thermal
  transport in tlinte2},\ }\href {https://doi.org/10.1038/s41524-020-00474-5}
  {\bibfield  {journal} {\bibinfo  {journal} {npj Computational Materials}\
  }\textbf {\bibinfo {volume} {7}},\ \bibinfo {pages} {5} (\bibinfo {year}
  {2021})}\BibitemShut {NoStop}%
\bibitem [{\citenamefont {Tamura}(1983)}]{Isotope_sctr}%
  \BibitemOpen
  \bibfield  {author} {\bibinfo {author} {\bibfnamefont {S.-i.}\ \bibnamefont
  {Tamura}},\ }\bibfield  {title} {\bibinfo {title} {Isotope scattering of
  dispersive phonons in ge},\ }\href {https://doi.org/10.1103/PhysRevB.27.858}
  {\bibfield  {journal} {\bibinfo  {journal} {Phys. Rev. B}\ }\textbf {\bibinfo
  {volume} {27}},\ \bibinfo {pages} {858} (\bibinfo {year} {1983})}\BibitemShut
  {NoStop}%
\bibitem [{\citenamefont {Ziman}(2001)}]{Ziman}%
  \BibitemOpen
  \bibfield  {author} {\bibinfo {author} {\bibfnamefont {J.}~\bibnamefont
  {Ziman}},\ }\href@noop {} {\emph {\bibinfo {title} {Electrons and Phonons:
  The Theory of Transport Phenomena in Solids}}}\ (\bibinfo  {publisher}
  {Oxford University Press},\ \bibinfo {year} {2001})\BibitemShut {NoStop}%
\bibitem [{\citenamefont {Broido}\ \emph {et~al.}(2005)\citenamefont {Broido},
  \citenamefont {Ward},\ and\ \citenamefont {Mingo}}]{BTE_lin_iter}%
  \BibitemOpen
  \bibfield  {author} {\bibinfo {author} {\bibfnamefont {D.~A.}\ \bibnamefont
  {Broido}}, \bibinfo {author} {\bibfnamefont {A.}~\bibnamefont {Ward}},\ and\
  \bibinfo {author} {\bibfnamefont {N.}~\bibnamefont {Mingo}},\ }\bibfield
  {title} {\bibinfo {title} {Lattice thermal conductivity of silicon from
  empirical interatomic potentials},\ }\href
  {https://doi.org/10.1103/PhysRevB.72.014308} {\bibfield  {journal} {\bibinfo
  {journal} {Phys. Rev. B}\ }\textbf {\bibinfo {volume} {72}},\ \bibinfo
  {pages} {014308} (\bibinfo {year} {2005})}\BibitemShut {NoStop}%
\bibitem [{\citenamefont {Lindsay}\ and\ \citenamefont
  {Broido}(2008)}]{Lindsay_2008}%
  \BibitemOpen
  \bibfield  {author} {\bibinfo {author} {\bibfnamefont {L.}~\bibnamefont
  {Lindsay}}\ and\ \bibinfo {author} {\bibfnamefont {D.~A.}\ \bibnamefont
  {Broido}},\ }\bibfield  {title} {\bibinfo {title} {Three-phonon phase space
  and lattice thermal conductivity in semiconductors},\ }\href
  {https://doi.org/10.1088/0953-8984/20/16/165209} {\bibfield  {journal}
  {\bibinfo  {journal} {Journal of Physics: Condensed Matter}\ }\textbf
  {\bibinfo {volume} {20}},\ \bibinfo {pages} {165209} (\bibinfo {year}
  {2008})}\BibitemShut {NoStop}%
\bibitem [{\citenamefont {Omini}\ and\ \citenamefont
  {Sparavigna}(1995)}]{itr_BTE}%
  \BibitemOpen
  \bibfield  {author} {\bibinfo {author} {\bibfnamefont {M.}~\bibnamefont
  {Omini}}\ and\ \bibinfo {author} {\bibfnamefont {A.}~\bibnamefont
  {Sparavigna}},\ }\bibfield  {title} {\bibinfo {title} {An iterative approach
  to the phonon boltzmann equation in the theory of thermal conductivity},\
  }\href {https://doi.org/https://doi.org/10.1016/0921-4526(95)00016-3}
  {\bibfield  {journal} {\bibinfo  {journal} {Physica B: Condensed Matter}\
  }\textbf {\bibinfo {volume} {212}},\ \bibinfo {pages} {101} (\bibinfo {year}
  {1995})}\BibitemShut {NoStop}%
\bibitem [{\citenamefont {MacDonald}\ \emph {et~al.}(1979)\citenamefont
  {MacDonald}, \citenamefont {Vosko},\ and\ \citenamefont
  {Coleridge}}]{Tetrahedron}%
  \BibitemOpen
  \bibfield  {author} {\bibinfo {author} {\bibfnamefont {A.~H.}\ \bibnamefont
  {MacDonald}}, \bibinfo {author} {\bibfnamefont {S.~H.}\ \bibnamefont
  {Vosko}},\ and\ \bibinfo {author} {\bibfnamefont {P.~T.}\ \bibnamefont
  {Coleridge}},\ }\bibfield  {title} {\bibinfo {title} {Extensions of the
  tetrahedron method for evaluating spectral properties of solids},\ }\href
  {https://doi.org/10.1088/0022-3719/12/15/008} {\bibfield  {journal} {\bibinfo
   {journal} {Journal of Physics C: Solid State Physics}\ }\textbf {\bibinfo
  {volume} {12}},\ \bibinfo {pages} {2991} (\bibinfo {year}
  {1979})}\BibitemShut {NoStop}%
\bibitem [{\citenamefont {Perdew}\ \emph {et~al.}(1996)\citenamefont {Perdew},
  \citenamefont {Burke},\ and\ \citenamefont {Ernzerhof}}]{GGA-PBE}%
  \BibitemOpen
  \bibfield  {author} {\bibinfo {author} {\bibfnamefont {J.~P.}\ \bibnamefont
  {Perdew}}, \bibinfo {author} {\bibfnamefont {K.}~\bibnamefont {Burke}},\ and\
  \bibinfo {author} {\bibfnamefont {M.}~\bibnamefont {Ernzerhof}},\ }\bibfield
  {title} {\bibinfo {title} {Generalized gradient approximation made simple},\
  }\href {https://doi.org/10.1103/PhysRevLett.77.3865} {\bibfield  {journal}
  {\bibinfo  {journal} {Phys. Rev. Lett.}\ }\textbf {\bibinfo {volume} {77}},\
  \bibinfo {pages} {3865} (\bibinfo {year} {1996})}\BibitemShut {NoStop}%
\bibitem [{\citenamefont {Sanjurjo}\ \emph {et~al.}(1983)\citenamefont
  {Sanjurjo}, \citenamefont {L\'opez-Cruz}, \citenamefont {Vogl},\ and\
  \citenamefont {Cardona}}]{BN_phonon_disp_Exp1}%
  \BibitemOpen
  \bibfield  {author} {\bibinfo {author} {\bibfnamefont {J.~A.}\ \bibnamefont
  {Sanjurjo}}, \bibinfo {author} {\bibfnamefont {E.}~\bibnamefont
  {L\'opez-Cruz}}, \bibinfo {author} {\bibfnamefont {P.}~\bibnamefont {Vogl}},\
  and\ \bibinfo {author} {\bibfnamefont {M.}~\bibnamefont {Cardona}},\
  }\bibfield  {title} {\bibinfo {title} {Dependence on volume of the phonon
  frequencies and the ir effective charges of several iii-v semiconductors},\
  }\href {https://doi.org/10.1103/PhysRevB.28.4579} {\bibfield  {journal}
  {\bibinfo  {journal} {Phys. Rev. B}\ }\textbf {\bibinfo {volume} {28}},\
  \bibinfo {pages} {4579} (\bibinfo {year} {1983})}\BibitemShut {NoStop}%
\bibitem [{\citenamefont {Reich}\ \emph {et~al.}(2005)\citenamefont {Reich},
  \citenamefont {Ferrari}, \citenamefont {Arenal}, \citenamefont {Loiseau},
  \citenamefont {Bello},\ and\ \citenamefont
  {Robertson}}]{BN_phonon_disp_Exp2}%
  \BibitemOpen
  \bibfield  {author} {\bibinfo {author} {\bibfnamefont {S.}~\bibnamefont
  {Reich}}, \bibinfo {author} {\bibfnamefont {A.~C.}\ \bibnamefont {Ferrari}},
  \bibinfo {author} {\bibfnamefont {R.}~\bibnamefont {Arenal}}, \bibinfo
  {author} {\bibfnamefont {A.}~\bibnamefont {Loiseau}}, \bibinfo {author}
  {\bibfnamefont {I.}~\bibnamefont {Bello}},\ and\ \bibinfo {author}
  {\bibfnamefont {J.}~\bibnamefont {Robertson}},\ }\bibfield  {title} {\bibinfo
  {title} {Resonant raman scattering in cubic and hexagonal boron nitride},\
  }\href {https://doi.org/10.1103/PhysRevB.71.205201} {\bibfield  {journal}
  {\bibinfo  {journal} {Phys. Rev. B}\ }\textbf {\bibinfo {volume} {71}},\
  \bibinfo {pages} {205201} (\bibinfo {year} {2005})}\BibitemShut {NoStop}%
\bibitem [{\citenamefont {Kikkawa}\ \emph {et~al.}(2021)\citenamefont
  {Kikkawa}, \citenamefont {Taniguchi},\ and\ \citenamefont
  {Kimoto}}]{BN_phonon_disp_Exp3}%
  \BibitemOpen
  \bibfield  {author} {\bibinfo {author} {\bibfnamefont {J.}~\bibnamefont
  {Kikkawa}}, \bibinfo {author} {\bibfnamefont {T.}~\bibnamefont {Taniguchi}},\
  and\ \bibinfo {author} {\bibfnamefont {K.}~\bibnamefont {Kimoto}},\
  }\bibfield  {title} {\bibinfo {title} {Nanometric phonon spectroscopy for
  diamond and cubic boron nitride},\ }\href
  {https://doi.org/10.1103/PhysRevB.104.L201402} {\bibfield  {journal}
  {\bibinfo  {journal} {Phys. Rev. B}\ }\textbf {\bibinfo {volume} {104}},\
  \bibinfo {pages} {L201402} (\bibinfo {year} {2021})}\BibitemShut {NoStop}%
\bibitem [{\citenamefont {Chen}\ \emph {et~al.}(2020)\citenamefont {Chen},
  \citenamefont {Song}, \citenamefont {Ravichandran}, \citenamefont {Zheng},
  \citenamefont {Chen}, \citenamefont {Lee}, \citenamefont {Sun}, \citenamefont
  {Li}, \citenamefont {Gamage}, \citenamefont {Tian}, \citenamefont {Ding},
  \citenamefont {Song}, \citenamefont {Rai}, \citenamefont {Wu}, \citenamefont
  {Koirala}, \citenamefont {Schmidt}, \citenamefont {Watanabe}, \citenamefont
  {Lv}, \citenamefont {Ren}, \citenamefont {Shi}, \citenamefont {Cahill},
  \citenamefont {Taniguchi}, \citenamefont {Broido},\ and\ \citenamefont
  {Chen}}]{Exp_kappa_2020}%
  \BibitemOpen
  \bibfield  {author} {\bibinfo {author} {\bibfnamefont {K.}~\bibnamefont
  {Chen}}, \bibinfo {author} {\bibfnamefont {B.}~\bibnamefont {Song}}, \bibinfo
  {author} {\bibfnamefont {N.~K.}\ \bibnamefont {Ravichandran}}, \bibinfo
  {author} {\bibfnamefont {Q.}~\bibnamefont {Zheng}}, \bibinfo {author}
  {\bibfnamefont {X.}~\bibnamefont {Chen}}, \bibinfo {author} {\bibfnamefont
  {H.}~\bibnamefont {Lee}}, \bibinfo {author} {\bibfnamefont {H.}~\bibnamefont
  {Sun}}, \bibinfo {author} {\bibfnamefont {S.}~\bibnamefont {Li}}, \bibinfo
  {author} {\bibfnamefont {G.~A. G.~U.}\ \bibnamefont {Gamage}}, \bibinfo
  {author} {\bibfnamefont {F.}~\bibnamefont {Tian}}, \bibinfo {author}
  {\bibfnamefont {Z.}~\bibnamefont {Ding}}, \bibinfo {author} {\bibfnamefont
  {Q.}~\bibnamefont {Song}}, \bibinfo {author} {\bibfnamefont {A.}~\bibnamefont
  {Rai}}, \bibinfo {author} {\bibfnamefont {H.}~\bibnamefont {Wu}}, \bibinfo
  {author} {\bibfnamefont {P.}~\bibnamefont {Koirala}}, \bibinfo {author}
  {\bibfnamefont {A.~J.}\ \bibnamefont {Schmidt}}, \bibinfo {author}
  {\bibfnamefont {K.}~\bibnamefont {Watanabe}}, \bibinfo {author}
  {\bibfnamefont {B.}~\bibnamefont {Lv}}, \bibinfo {author} {\bibfnamefont
  {Z.}~\bibnamefont {Ren}}, \bibinfo {author} {\bibfnamefont {L.}~\bibnamefont
  {Shi}}, \bibinfo {author} {\bibfnamefont {D.~G.}\ \bibnamefont {Cahill}},
  \bibinfo {author} {\bibfnamefont {T.}~\bibnamefont {Taniguchi}}, \bibinfo
  {author} {\bibfnamefont {D.}~\bibnamefont {Broido}},\ and\ \bibinfo {author}
  {\bibfnamefont {G.}~\bibnamefont {Chen}},\ }\bibfield  {title} {\bibinfo
  {title} {Ultrahigh thermal conductivity in isotope-enriched cubic boron
  nitride},\ }\href {https://doi.org/10.1126/science.aaz6149} {\bibfield
  {journal} {\bibinfo  {journal} {Science}\ }\textbf {\bibinfo {volume}
  {367}},\ \bibinfo {pages} {555} (\bibinfo {year} {2020})},\ \Eprint
  {https://arxiv.org/abs/https://www.science.org/doi/pdf/10.1126/science.aaz6149}
  {https://www.science.org/doi/pdf/10.1126/science.aaz6149} \BibitemShut
  {NoStop}%
\bibitem [{\citenamefont {Kang}\ \emph {et~al.}(2018)\citenamefont {Kang},
  \citenamefont {Li}, \citenamefont {Wu}, \citenamefont {Nguyen},\ and\
  \citenamefont {Hu}}]{Exp_kappa_2018}%
  \BibitemOpen
  \bibfield  {author} {\bibinfo {author} {\bibfnamefont {J.~S.}\ \bibnamefont
  {Kang}}, \bibinfo {author} {\bibfnamefont {M.}~\bibnamefont {Li}}, \bibinfo
  {author} {\bibfnamefont {H.}~\bibnamefont {Wu}}, \bibinfo {author}
  {\bibfnamefont {H.}~\bibnamefont {Nguyen}},\ and\ \bibinfo {author}
  {\bibfnamefont {Y.}~\bibnamefont {Hu}},\ }\bibfield  {title} {\bibinfo
  {title} {Experimental observation of high thermal conductivity in boron
  arsenide},\ }\href {https://doi.org/10.1126/science.aat5522} {\bibfield
  {journal} {\bibinfo  {journal} {Science}\ }\textbf {\bibinfo {volume}
  {361}},\ \bibinfo {pages} {575} (\bibinfo {year} {2018})},\ \Eprint
  {https://arxiv.org/abs/https://www.science.org/doi/pdf/10.1126/science.aat5522}
  {https://www.science.org/doi/pdf/10.1126/science.aat5522} \BibitemShut
  {NoStop}%
\bibitem [{\citenamefont {Serrano}\ \emph {et~al.}(2002)\citenamefont
  {Serrano}, \citenamefont {Strempfer}, \citenamefont {Cardona}, \citenamefont
  {Schwoerer-Böhning}, \citenamefont {Requardt}, \citenamefont {Lorenzen},
  \citenamefont {Stojetz}, \citenamefont {Pavone},\ and\ \citenamefont
  {Choyke}}]{SiC_phonon_disp}%
  \BibitemOpen
  \bibfield  {author} {\bibinfo {author} {\bibfnamefont {J.}~\bibnamefont
  {Serrano}}, \bibinfo {author} {\bibfnamefont {J.}~\bibnamefont {Strempfer}},
  \bibinfo {author} {\bibfnamefont {M.}~\bibnamefont {Cardona}}, \bibinfo
  {author} {\bibfnamefont {M.}~\bibnamefont {Schwoerer-Böhning}}, \bibinfo
  {author} {\bibfnamefont {H.}~\bibnamefont {Requardt}}, \bibinfo {author}
  {\bibfnamefont {M.}~\bibnamefont {Lorenzen}}, \bibinfo {author}
  {\bibfnamefont {B.}~\bibnamefont {Stojetz}}, \bibinfo {author} {\bibfnamefont
  {P.}~\bibnamefont {Pavone}},\ and\ \bibinfo {author} {\bibfnamefont {W.~J.}\
  \bibnamefont {Choyke}},\ }\bibfield  {title} {\bibinfo {title}
  {{Determination of the phonon dispersion of zinc blende (3C) silicon carbide
  by inelastic x-ray scattering}},\ }\href {https://doi.org/10.1063/1.1484241}
  {\bibfield  {journal} {\bibinfo  {journal} {Applied Physics Letters}\
  }\textbf {\bibinfo {volume} {80}},\ \bibinfo {pages} {4360} (\bibinfo {year}
  {2002})},\ \Eprint
  {https://arxiv.org/abs/https://pubs.aip.org/aip/apl/article-pdf/80/23/4360/10192323/4360\_1\_online.pdf}
  {https://pubs.aip.org/aip/apl/article-pdf/80/23/4360/10192323/4360\_1\_online.pdf}
  \BibitemShut {NoStop}%
\bibitem [{\citenamefont {Morelli}\ \emph {et~al.}(2002)\citenamefont
  {Morelli}, \citenamefont {Heremans},\ and\ \citenamefont
  {Slack}}]{kappa_singleCrystallineSiC}%
  \BibitemOpen
  \bibfield  {author} {\bibinfo {author} {\bibfnamefont {D.~T.}\ \bibnamefont
  {Morelli}}, \bibinfo {author} {\bibfnamefont {J.~P.}\ \bibnamefont
  {Heremans}},\ and\ \bibinfo {author} {\bibfnamefont {G.~A.}\ \bibnamefont
  {Slack}},\ }\bibfield  {title} {\bibinfo {title} {Estimation of the isotope
  effect on the lattice thermal conductivity of group iv and group iii-v
  semiconductors},\ }\href {https://doi.org/10.1103/PhysRevB.66.195304}
  {\bibfield  {journal} {\bibinfo  {journal} {Phys. Rev. B}\ }\textbf {\bibinfo
  {volume} {66}},\ \bibinfo {pages} {195304} (\bibinfo {year}
  {2002})}\BibitemShut {NoStop}%
\bibitem [{\citenamefont {Ivanova}\ \emph {et~al.}(2006)\citenamefont
  {Ivanova}, \citenamefont {Aleksandrov},\ and\ \citenamefont
  {Demakov}}]{kappa_polyCrystallineSiC}%
  \BibitemOpen
  \bibfield  {author} {\bibinfo {author} {\bibfnamefont {L.~M.}\ \bibnamefont
  {Ivanova}}, \bibinfo {author} {\bibfnamefont {P.~A.}\ \bibnamefont
  {Aleksandrov}},\ and\ \bibinfo {author} {\bibfnamefont {K.~D.}\ \bibnamefont
  {Demakov}},\ }\bibfield  {title} {\bibinfo {title} {Thermoelectric properties
  of vapor-grown polycrystalline cubic sic},\ }\href
  {https://doi.org/10.1134/S0020168506110069} {\bibfield  {journal} {\bibinfo
  {journal} {Inorganic Materials}\ }\textbf {\bibinfo {volume} {42}},\ \bibinfo
  {pages} {1205} (\bibinfo {year} {2006})}\BibitemShut {NoStop}%
\bibitem [{\citenamefont {Raunio}\ \emph {et~al.}(1969)\citenamefont {Raunio},
  \citenamefont {Almqvist},\ and\ \citenamefont {Stedman}}]{Raunio_PhononDisp}%
  \BibitemOpen
  \bibfield  {author} {\bibinfo {author} {\bibfnamefont {G.}~\bibnamefont
  {Raunio}}, \bibinfo {author} {\bibfnamefont {L.}~\bibnamefont {Almqvist}},\
  and\ \bibinfo {author} {\bibfnamefont {R.}~\bibnamefont {Stedman}},\
  }\bibfield  {title} {\bibinfo {title} {Phonon dispersion relations in nacl},\
  }\href {https://doi.org/10.1103/PhysRev.178.1496} {\bibfield  {journal}
  {\bibinfo  {journal} {Phys. Rev.}\ }\textbf {\bibinfo {volume} {178}},\
  \bibinfo {pages} {1496} (\bibinfo {year} {1969})}\BibitemShut {NoStop}%
\bibitem [{\citenamefont {Cowley}\ \emph {et~al.}(1983)\citenamefont {Cowley},
  \citenamefont {Satija},\ and\ \citenamefont {Youngblood}}]{Linewidth}%
  \BibitemOpen
  \bibfield  {author} {\bibinfo {author} {\bibfnamefont {E.~R.}\ \bibnamefont
  {Cowley}}, \bibinfo {author} {\bibfnamefont {S.}~\bibnamefont {Satija}},\
  and\ \bibinfo {author} {\bibfnamefont {R.}~\bibnamefont {Youngblood}},\
  }\bibfield  {title} {\bibinfo {title} {Line shapes of longitudinal-optic
  phonons in sodium chloride at 300 and 600 k},\ }\href
  {https://doi.org/10.1103/PhysRevB.28.993} {\bibfield  {journal} {\bibinfo
  {journal} {Phys. Rev. B}\ }\textbf {\bibinfo {volume} {28}},\ \bibinfo
  {pages} {993} (\bibinfo {year} {1983})}\BibitemShut {NoStop}%
\bibitem [{\citenamefont {Håkansson}\ and\ \citenamefont
  {Andersson}(1986)}]{HAKANSSON_Exp_Pressure}%
  \BibitemOpen
  \bibfield  {author} {\bibinfo {author} {\bibfnamefont {B.}~\bibnamefont
  {Håkansson}}\ and\ \bibinfo {author} {\bibfnamefont {P.}~\bibnamefont
  {Andersson}},\ }\bibfield  {title} {\bibinfo {title} {Thermal conductivity
  and heat capacity of solid nacl and nai under pressure},\ }\href
  {https://doi.org/https://doi.org/10.1016/0022-3697(86)90025-9} {\bibfield
  {journal} {\bibinfo  {journal} {Journal of Physics and Chemistry of Solids}\
  }\textbf {\bibinfo {volume} {47}},\ \bibinfo {pages} {355} (\bibinfo {year}
  {1986})}\BibitemShut {NoStop}%
\bibitem [{\citenamefont {Klein}\ and\ \citenamefont
  {Caldwell}(2004)}]{Klein_Exp}%
  \BibitemOpen
  \bibfield  {author} {\bibinfo {author} {\bibfnamefont {M.~V.}\ \bibnamefont
  {Klein}}\ and\ \bibinfo {author} {\bibfnamefont {R.~F.}\ \bibnamefont
  {Caldwell}},\ }\bibfield  {title} {\bibinfo {title} {{Low Temperature System
  for Thermal Conductivity Measurements}},\ }\href
  {https://doi.org/10.1063/1.1719962} {\bibfield  {journal} {\bibinfo
  {journal} {Review of Scientific Instruments}\ }\textbf {\bibinfo {volume}
  {37}},\ \bibinfo {pages} {1291} (\bibinfo {year} {2004})},\ \Eprint
  {https://arxiv.org/abs/https://pubs.aip.org/aip/rsi/article-pdf/37/10/1291/8352456/1291\_1\_online.pdf}
  {https://pubs.aip.org/aip/rsi/article-pdf/37/10/1291/8352456/1291\_1\_online.pdf}
  \BibitemShut {NoStop}%
\bibitem [{\citenamefont {McCarthy}\ and\ \citenamefont
  {Ballard}(2004)}]{McCarthy_Exp}%
  \BibitemOpen
  \bibfield  {author} {\bibinfo {author} {\bibfnamefont {K.~A.}\ \bibnamefont
  {McCarthy}}\ and\ \bibinfo {author} {\bibfnamefont {S.~S.}\ \bibnamefont
  {Ballard}},\ }\bibfield  {title} {\bibinfo {title} {{Thermal Conductivity of
  Eight Halide Crystals in the Temperature Range 220°K to 390°K}},\ }\href
  {https://doi.org/10.1063/1.1735853} {\bibfield  {journal} {\bibinfo
  {journal} {Journal of Applied Physics}\ }\textbf {\bibinfo {volume} {31}},\
  \bibinfo {pages} {1410} (\bibinfo {year} {2004})},\ \Eprint
  {https://arxiv.org/abs/https://pubs.aip.org/aip/jap/article-pdf/31/8/1410/7928241/1410\_1\_online.pdf}
  {https://pubs.aip.org/aip/jap/article-pdf/31/8/1410/7928241/1410\_1\_online.pdf}
  \BibitemShut {NoStop}%
\bibitem [{\citenamefont {Goetz}\ and\ \citenamefont
  {Cowen}(1982)}]{AgI_Exp_1982}%
  \BibitemOpen
  \bibfield  {author} {\bibinfo {author} {\bibfnamefont {M.}~\bibnamefont
  {Goetz}}\ and\ \bibinfo {author} {\bibfnamefont {J.}~\bibnamefont {Cowen}},\
  }\bibfield  {title} {\bibinfo {title} {The thermal conductivity of silver
  iodide},\ }\href
  {https://doi.org/https://doi.org/10.1016/0038-1098(82)90377-5} {\bibfield
  {journal} {\bibinfo  {journal} {Solid State Communications}\ }\textbf
  {\bibinfo {volume} {41}},\ \bibinfo {pages} {293} (\bibinfo {year}
  {1982})}\BibitemShut {NoStop}%
\bibitem [{\citenamefont {Wang}\ \emph {et~al.}(2023)\citenamefont {Wang},
  \citenamefont {Gan}, \citenamefont {Hu}, \citenamefont {Li}, \citenamefont
  {Xie},\ and\ \citenamefont {He}}]{AgI_2023_kappa_1}%
  \BibitemOpen
  \bibfield  {author} {\bibinfo {author} {\bibfnamefont {Y.}~\bibnamefont
  {Wang}}, \bibinfo {author} {\bibfnamefont {Q.}~\bibnamefont {Gan}}, \bibinfo
  {author} {\bibfnamefont {M.}~\bibnamefont {Hu}}, \bibinfo {author}
  {\bibfnamefont {J.}~\bibnamefont {Li}}, \bibinfo {author} {\bibfnamefont
  {L.}~\bibnamefont {Xie}},\ and\ \bibinfo {author} {\bibfnamefont
  {J.}~\bibnamefont {He}},\ }\bibfield  {title} {\bibinfo {title} {Anharmonic
  lattice dynamics and the origin of intrinsic ultralow thermal conductivity in
  agi materials},\ }\href {https://doi.org/10.1103/PhysRevB.107.064308}
  {\bibfield  {journal} {\bibinfo  {journal} {Phys. Rev. B}\ }\textbf {\bibinfo
  {volume} {107}},\ \bibinfo {pages} {064308} (\bibinfo {year}
  {2023})}\BibitemShut {NoStop}%
\bibitem [{\citenamefont {Acharyya}\ \emph {et~al.}(2022)\citenamefont
  {Acharyya}, \citenamefont {Ghosh}, \citenamefont {Pal}, \citenamefont {Rana},
  \citenamefont {Dutta}, \citenamefont {Swain}, \citenamefont {Etter},
  \citenamefont {Soni}, \citenamefont {Waghmare},\ and\ \citenamefont
  {Biswas}}]{Kanishka2022}%
  \BibitemOpen
  \bibfield  {author} {\bibinfo {author} {\bibfnamefont {P.}~\bibnamefont
  {Acharyya}}, \bibinfo {author} {\bibfnamefont {T.}~\bibnamefont {Ghosh}},
  \bibinfo {author} {\bibfnamefont {K.}~\bibnamefont {Pal}}, \bibinfo {author}
  {\bibfnamefont {K.~S.}\ \bibnamefont {Rana}}, \bibinfo {author}
  {\bibfnamefont {M.}~\bibnamefont {Dutta}}, \bibinfo {author} {\bibfnamefont
  {D.}~\bibnamefont {Swain}}, \bibinfo {author} {\bibfnamefont
  {M.}~\bibnamefont {Etter}}, \bibinfo {author} {\bibfnamefont
  {A.}~\bibnamefont {Soni}}, \bibinfo {author} {\bibfnamefont {U.~V.}\
  \bibnamefont {Waghmare}},\ and\ \bibinfo {author} {\bibfnamefont
  {K.}~\bibnamefont {Biswas}},\ }\bibfield  {title} {\bibinfo {title} {Glassy
  thermal conductivity in cs3bi2i6cl3 single crystal},\ }\href
  {https://doi.org/10.1038/s41467-022-32773-4} {\bibfield  {journal} {\bibinfo
  {journal} {Nature Communications}\ }\textbf {\bibinfo {volume} {13}},\
  \bibinfo {pages} {5053} (\bibinfo {year} {2022})}\BibitemShut {NoStop}%
\bibitem [{\citenamefont {Acharyya}\ \emph {et~al.}(2023)\citenamefont
  {Acharyya}, \citenamefont {Pal}, \citenamefont {Ahad}, \citenamefont
  {Sarkar}, \citenamefont {Rana}, \citenamefont {Dutta}, \citenamefont {Soni},
  \citenamefont {Waghmare},\ and\ \citenamefont {Biswas}}]{Kanishka2023}%
  \BibitemOpen
  \bibfield  {author} {\bibinfo {author} {\bibfnamefont {P.}~\bibnamefont
  {Acharyya}}, \bibinfo {author} {\bibfnamefont {K.}~\bibnamefont {Pal}},
  \bibinfo {author} {\bibfnamefont {A.}~\bibnamefont {Ahad}}, \bibinfo {author}
  {\bibfnamefont {D.}~\bibnamefont {Sarkar}}, \bibinfo {author} {\bibfnamefont
  {K.~S.}\ \bibnamefont {Rana}}, \bibinfo {author} {\bibfnamefont
  {M.}~\bibnamefont {Dutta}}, \bibinfo {author} {\bibfnamefont
  {A.}~\bibnamefont {Soni}}, \bibinfo {author} {\bibfnamefont {U.~V.}\
  \bibnamefont {Waghmare}},\ and\ \bibinfo {author} {\bibfnamefont
  {K.}~\bibnamefont {Biswas}},\ }\bibfield  {title} {\bibinfo {title} {Extended
  antibonding states and phonon localization induce ultralow thermal
  conductivity in low dimensional metal halide},\ }\href
  {https://doi.org/https://doi.org/10.1002/adfm.202304607} {\bibfield
  {journal} {\bibinfo  {journal} {Advanced Functional Materials}\ }\textbf
  {\bibinfo {volume} {33}},\ \bibinfo {pages} {2304607} (\bibinfo {year}
  {2023})},\ \Eprint
  {https://arxiv.org/abs/https://onlinelibrary.wiley.com/doi/pdf/10.1002/adfm.202304607}
  {https://onlinelibrary.wiley.com/doi/pdf/10.1002/adfm.202304607} \BibitemShut
  {NoStop}%
\bibitem [{\citenamefont {Ouyang}\ \emph {et~al.}(2023)\citenamefont {Ouyang},
  \citenamefont {Zeng}, \citenamefont {Wang}, \citenamefont {Wang},\ and\
  \citenamefont {Chen}}]{AgI_2023_kappa_2}%
  \BibitemOpen
  \bibfield  {author} {\bibinfo {author} {\bibfnamefont {N.}~\bibnamefont
  {Ouyang}}, \bibinfo {author} {\bibfnamefont {Z.}~\bibnamefont {Zeng}},
  \bibinfo {author} {\bibfnamefont {C.}~\bibnamefont {Wang}}, \bibinfo {author}
  {\bibfnamefont {Q.}~\bibnamefont {Wang}},\ and\ \bibinfo {author}
  {\bibfnamefont {Y.}~\bibnamefont {Chen}},\ }\bibfield  {title} {\bibinfo
  {title} {Role of high-order lattice anharmonicity in the phonon thermal
  transport of silver halide $\mathrm{Ag}x
  (x=\mathrm{Cl},\mathrm{Br},\mathrm{I})$},\ }\href
  {https://doi.org/10.1103/PhysRevB.108.174302} {\bibfield  {journal} {\bibinfo
   {journal} {Phys. Rev. B}\ }\textbf {\bibinfo {volume} {108}},\ \bibinfo
  {pages} {174302} (\bibinfo {year} {2023})}\BibitemShut {NoStop}%
\bibitem [{\citenamefont {Hsieh}(2021)}]{Hsieh2021_Pressure}%
  \BibitemOpen
  \bibfield  {author} {\bibinfo {author} {\bibfnamefont {W.-P.}\ \bibnamefont
  {Hsieh}},\ }\bibfield  {title} {\bibinfo {title} {High-pressure thermal
  conductivity and compressional velocity of nacl in b1 and b2 phase},\ }\href
  {https://doi.org/10.1038/s41598-021-00736-2} {\bibfield  {journal} {\bibinfo
  {journal} {Scientific Reports}\ }\textbf {\bibinfo {volume} {11}},\ \bibinfo
  {pages} {21321} (\bibinfo {year} {2021})}\BibitemShut {NoStop}%
\bibitem [{\citenamefont {Srivastava}\ and\ \citenamefont
  {Merchant}(1973)}]{thermalExpn}%
  \BibitemOpen
  \bibfield  {author} {\bibinfo {author} {\bibfnamefont {K.}~\bibnamefont
  {Srivastava}}\ and\ \bibinfo {author} {\bibfnamefont {H.}~\bibnamefont
  {Merchant}},\ }\bibfield  {title} {\bibinfo {title} {Thermal expansion of
  alkali halides above 300°k},\ }\href
  {https://doi.org/https://doi.org/10.1016/S0022-3697(73)80055-1} {\bibfield
  {journal} {\bibinfo  {journal} {Journal of Physics and Chemistry of Solids}\
  }\textbf {\bibinfo {volume} {34}},\ \bibinfo {pages} {2069} (\bibinfo {year}
  {1973})}\BibitemShut {NoStop}%
\bibitem [{\citenamefont {Ravichandran}\ and\ \citenamefont
  {Broido}(2019)}]{Ravichandran2019}%
  \BibitemOpen
  \bibfield  {author} {\bibinfo {author} {\bibfnamefont {N.~K.}\ \bibnamefont
  {Ravichandran}}\ and\ \bibinfo {author} {\bibfnamefont {D.}~\bibnamefont
  {Broido}},\ }\bibfield  {title} {\bibinfo {title} {Non-monotonic pressure
  dependence of the thermal conductivity of boron arsenide},\ }\href
  {https://doi.org/10.1038/s41467-019-08713-0} {\bibfield  {journal} {\bibinfo
  {journal} {Nature Communications}\ }\textbf {\bibinfo {volume} {10}},\
  \bibinfo {pages} {827} (\bibinfo {year} {2019})}\BibitemShut {NoStop}%
\bibitem [{\citenamefont {Lindsay}\ \emph {et~al.}(2013)\citenamefont
  {Lindsay}, \citenamefont {Broido},\ and\ \citenamefont
  {Reinecke}}]{LinsdeyPhaseSpace}%
  \BibitemOpen
  \bibfield  {author} {\bibinfo {author} {\bibfnamefont {L.}~\bibnamefont
  {Lindsay}}, \bibinfo {author} {\bibfnamefont {D.~A.}\ \bibnamefont
  {Broido}},\ and\ \bibinfo {author} {\bibfnamefont {T.~L.}\ \bibnamefont
  {Reinecke}},\ }\bibfield  {title} {\bibinfo {title} {First-principles
  determination of ultrahigh thermal conductivity of boron arsenide: A
  competitor for diamond?},\ }\href
  {https://doi.org/10.1103/PhysRevLett.111.025901} {\bibfield  {journal}
  {\bibinfo  {journal} {Phys. Rev. Lett.}\ }\textbf {\bibinfo {volume} {111}},\
  \bibinfo {pages} {025901} (\bibinfo {year} {2013})}\BibitemShut {NoStop}%
\bibitem [{\citenamefont {Ravichandran}\ and\ \citenamefont
  {Broido}(2020)}]{NavneethaPRX}%
  \BibitemOpen
  \bibfield  {author} {\bibinfo {author} {\bibfnamefont {N.~K.}\ \bibnamefont
  {Ravichandran}}\ and\ \bibinfo {author} {\bibfnamefont {D.}~\bibnamefont
  {Broido}},\ }\bibfield  {title} {\bibinfo {title} {Phonon-phonon interactions
  in strongly bonded solids: Selection rules and higher-order processes},\
  }\href {https://doi.org/10.1103/PhysRevX.10.021063} {\bibfield  {journal}
  {\bibinfo  {journal} {Phys. Rev. X}\ }\textbf {\bibinfo {volume} {10}},\
  \bibinfo {pages} {021063} (\bibinfo {year} {2020})}\BibitemShut {NoStop}%
\bibitem [{\citenamefont {Protik}\ \emph {et~al.}(2022)\citenamefont {Protik},
  \citenamefont {Li}, \citenamefont {Pruneda}, \citenamefont {Broido},\ and\
  \citenamefont {Ordej{\'o}n}}]{Protik2022-eph}%
  \BibitemOpen
  \bibfield  {author} {\bibinfo {author} {\bibfnamefont {N.~H.}\ \bibnamefont
  {Protik}}, \bibinfo {author} {\bibfnamefont {C.}~\bibnamefont {Li}}, \bibinfo
  {author} {\bibfnamefont {M.}~\bibnamefont {Pruneda}}, \bibinfo {author}
  {\bibfnamefont {D.}~\bibnamefont {Broido}},\ and\ \bibinfo {author}
  {\bibfnamefont {P.}~\bibnamefont {Ordej{\'o}n}},\ }\bibfield  {title}
  {\bibinfo {title} {The elphbolt ab initio solver for the coupled
  electron-phonon boltzmann transport equations},\ }\href
  {https://doi.org/10.1038/s41524-022-00710-0} {\bibfield  {journal} {\bibinfo
  {journal} {npj Computational Materials}\ }\textbf {\bibinfo {volume} {8}},\
  \bibinfo {pages} {28} (\bibinfo {year} {2022})}\BibitemShut {NoStop}%
\end{thebibliography}%

\end{document}